\documentclass[fleqn,usenatbib]{mnras}

\usepackage{newtxtext,newtxmath}

\usepackage[T1]{fontenc}

\DeclareRobustCommand{\VAN}[3]{#2}
\let\VANthebibliography\thebibliography
\def\thebibliography{\DeclareRobustCommand{\VAN}[3]{##3}\VANthebibliography}

\usepackage{graphicx}	
\usepackage{amsmath}	
\usepackage{caption}
\usepackage{subcaption}

\newcommand{\HII}{\mbox{H\,{\footnotesize II}}}       
\newcommand{\reply}[1]{{#1}}

\title[A brown dwarf dust trail]{A possible trail of dust from a young, highly-extincted brown dwarf in the outskirts of the Trapezium Cluster}

\author[T. J. Haworth et al.]{
Thomas J. Haworth$^1$, Mark J. McCaughrean$^2$, Samuel G. Pearson$^3$ and Richard A. Booth$^4$
\\
$^{1}$Astronomy Unit, School of Physics and Astronomy, Queen Mary University of London, London E1 4NS, UK \\
$^{2}$Max-Planck-Institut f\"ur Astronomie, K\"onigstuhl~17, 69117 Heidelberg, DE \\
$^{3}$European Space Research and Technology Centre (ESTEC), European Space Agency, Postbus 299, 2200AG Noordwijk, NL \\
$^{4}$School of Physics and Astronomy, University of Leeds, Leeds LS2 9JT, UK
}

\date{Accepted XXX. Received YYY; in original form ZZZ}

\pubyear{2024}

\begin{document}
\label{firstpage}
\pagerange{\pageref{firstpage}--\pageref{lastpage}}
\maketitle

\begin{abstract}
We present the JWST discovery of a highly-extincted ($A_V\sim52$) candidate brown dwarf ($\sim0.018$M$_\odot$) in the outskirts of the Trapezium Cluster that appears to be coincident with the end of a $\sim 1700\,$au long, remarkably uniformly wide, dark trail that broadens only slightly at the end opposite the point source. We examine whether a dusty trail associated with a highly-extincted brown dwarf could plausibly be detected with JWST and explore possible origins. We show that a dusty trail associated with the brown dwarf could be observable if dust within it is larger than that in the ambient molecular cloud. For example, if the ambient cloud has a standard $\sim0.25$\,$\mu$m maximum grain size and the trail contains micron-sized grains, then the trail will have a scattering opacity over an order of magnitude larger compared to the surroundings in NIRCam short-wavelength filters. We use a simple model to show that a change in maximum grain size can reproduce the high $A_V$ and the multi-filter NIRCam contrast seen between the trail and its surroundings. We propose and explore two possible mechanisms that could be responsible for the trail: i) a weak FUV radiation-driven wind from the circum-brown dwarf disc due to the O stars in the region and ii) a Bondi-Hoyle-Lyttleton accretion wake. The former would be the most distant known case of the Trapezium stars' radiation driving winds from a disc, and the latter would be the first known example of ``late'' infall from the interstellar medium onto a low mass object in a high-mass star-forming region. 
\end{abstract}




\section{Introduction}
The Trapezium Cluster, in the inner part of the Orion Nebula Cluster (ONC), is famous as an environment where protoplanetary disc evolution (and by extension potentially planet formation) is significantly affected by the environment. We directly see teardrop-shaped plumes of material emanating from young star and disc systems in the central part of the Trapezium Cluster due to UV irradiation by the massive stars, most notably the O~stars $\theta^1$\,Ori C and $\theta^2$\,Ori A \citep{1994ApJ...436..194O, 2000AJ....119.2919B, 2008AJ....136.2136R}. For those teardrop-shaped ``proplyd'' systems, the mass-loss rate estimates are so high relative to the disc mass that the disc is expected to be dispersed on  timescales of order $10^4-10^5$\,years \citep[e.g.][]{1999AJ....118.2350H}, much shorter than the $10^6-10^7$\,year lifetimes typical of quiet regions \citep[e.g.][]{2001ApJ...553L.153H,2010A&A...510A..72F}. These high mass-loss rates are supported by theoretical models of ``external photoevaporation'' of discs \citep{2004ApJ...611..360A, 2016MNRAS.457.3593F, 2018MNRAS.481..452H, 2023MNRAS.526.4315H, 2022EPJP..137.1132W}. In recent years, attention has turned to the possible impact that external photoevaporation could have on planet populations and the role environment might play in contributing to exoplanet diversity. These studies do find that environmental irradiation can play a significant role in affecting disc evolution  \citep{2020MNRAS.492.1279S,2022MNRAS.514.2315C} and planet formation \citep[][]{2018MNRAS.474..886N, 2021A&A...656A..72B, 2022MNRAS.515.4287W, 2023MNRAS.522.1939Q, 2023EPJP..138..181E, 2024arXiv240719018H, 2024arXiv240809319H}. But environmental impact is not just limited to radiation \citep[e.g.][]{2022EPJP..137.1071R}: it also includes gravitational encounters \citep{1993MNRAS.261..190C, 2018ApJ...859..150R, 2023EPJP..138...11C} and ongoing accretion onto the system from the interstellar medium \citep{2020NatAs...4.1158P, 2023ApJ...953..190K, 2023ApJ...958...98F, 2024arXiv240110403G, 2024arXiv240917220W}. In particular, extended periods of accretion has recently been proposed as a potentially key driver of disc evolution \citep{2023EPJP..138..272K, 2024arXiv240507334P, 2024arXiv240508451W}. However, ongoing accretion has mostly been studied in low-mass star-forming regions where there is little stellar feedback and the UV radiation field is relatively weak. It will be important to understand the role it plays in the more massive star-forming regions, where stellar feedback disperses the molecular cloud and discs, because these environments are where a substantial fraction of stars form \citep[e.g.][]{2003ARA&A..41...57L,2020MNRAS.491..903W, 2022EPJP..137.1132W}.

A key question in the context of external photoevaporation is how pervasive its impact is. For example, how weak can the external UV radiation be while still driving significant mass loss, and what is the distribution and timescale that discs are exposed to certain levels of external irradiation? The proplyds in the Trapezium Cluster are subject to very strong FUV radiation fields, typically in the range $10^5-10^7$\,G$_0$\footnote{G$_0$ is the Habing unit of the radiation field, which broadly is normalised such that a value of 1G$_0$ is representative of the FUV in the interstellar medium of the Solar neighbourhood. Formally it is $1.6\times10^{-3}$\,erg\,cm$^{-2}$\,s$^{-1}$ integrated over the range 912-2400\AA. }. \cite{2008ApJ...675.1361F} computed theoretical distributions of FUV fields that stars will be exposed to. Ignoring extinction, this resulted in a distribution peaking at $\sim10^3$\,G$_0$ \citep[and we note that proplyds have been observed in such an environment][]{2016ApJ...826L..15K}. \cite{2020MNRAS.491..903W} undertook a similar exercise, predicting that external photoevaporation is prevalent. More recently \cite{2022MNRAS.512.3788Q} and \cite{2023MNRAS.520.5331W} directly computed the radiation field that stars are exposed to in star formation and feedback simulations, studying the time for which discs can be shielded from external irradiation by being embedded in their natal gas. So progress is being made addressing the ``when are discs irradiated'' and ``by how much FUV radiation'' questions associated with external photoevaporation; however, these do not address the question of how weak an environmental UV field can be while remaining significant.  

\cite{2023A&A...673L...2V} recently made a millimetre survey of protoplanetary discs across the wider Orion star-forming complex, finding statistical evidence that external irradiation can affect disc masses over a very wide range of UV fields, with a gradient in median dust masses down to around $\sim1$G$_0$. There is also some evidence for external winds from very extended discs in similarly low UV environments \citep{2017MNRAS.468L.108H}, though that is not conclusive \citep{2023ApJ...959L..27L}. In the Trapezium Cluster, our focus has predominantly been on the proplyds and the dramatic effect that external photoevaporation has on those. But slightly further afield and less studied, in the outskirts of the Trapezium Cluster, there are still many discs exposed to UV fields that are both predicted to result in non-negligible mass loss rate  \citep{2023MNRAS.526.4315H} and indirectly observed to impact disc masses by \cite{2023A&A...673L...2V}. Studying the outskirts of the Trapezium Cluster might provide new insight into external photoevaporation at weak/intermediate UV fields. With its high resolution and sensitivity, JWST offers the power to achieve this. 

The heart of the Orion Nebula and the Trapezium Cluster have been subject to recent study with JWST\@. The PDRs4All team imaged the Orion Bar (a photodissociation region) and some of the surrounding nebula \citep{2022PASP..134e4301B}, and simultaneously obtained data for a number of externally photoevaporating discs, most notably 203-506 that has been revealed to host a bounty of astrochemical discoveries \citep{2023arXiv231214056Z, 2023A&A...680A..19C, 2023Natur.621...56B, 2023MNRAS.525.4129H, 2024Sci...383..988P}. \cite{2023arXiv231003552M} undertook a much wider imaging study of the inner Orion Nebula and the Trapezium Cluster, surveying the dust, molecules, and ionised gas of the \HII{} region and the background molecular cloud, jets and outflows from embedded protostars, proplyds and silhouette discs, and the population of more than 3000 stars, brown dwarfs, and planetary-mass objects spanning the initial mass function. 

These latter observations of the inner Orion Nebula cover a field $11\times7.5$ arcmin providing a high-sensitivity, high-resolution infrared view of stars and discs both close to the central OB stars are up to 0.5\,pc further out, where intermediate-to-low UV irradiation may still be affecting disc evolution. 

\begin{figure*}
    \centering
    \includegraphics[width=1.8\columnwidth]{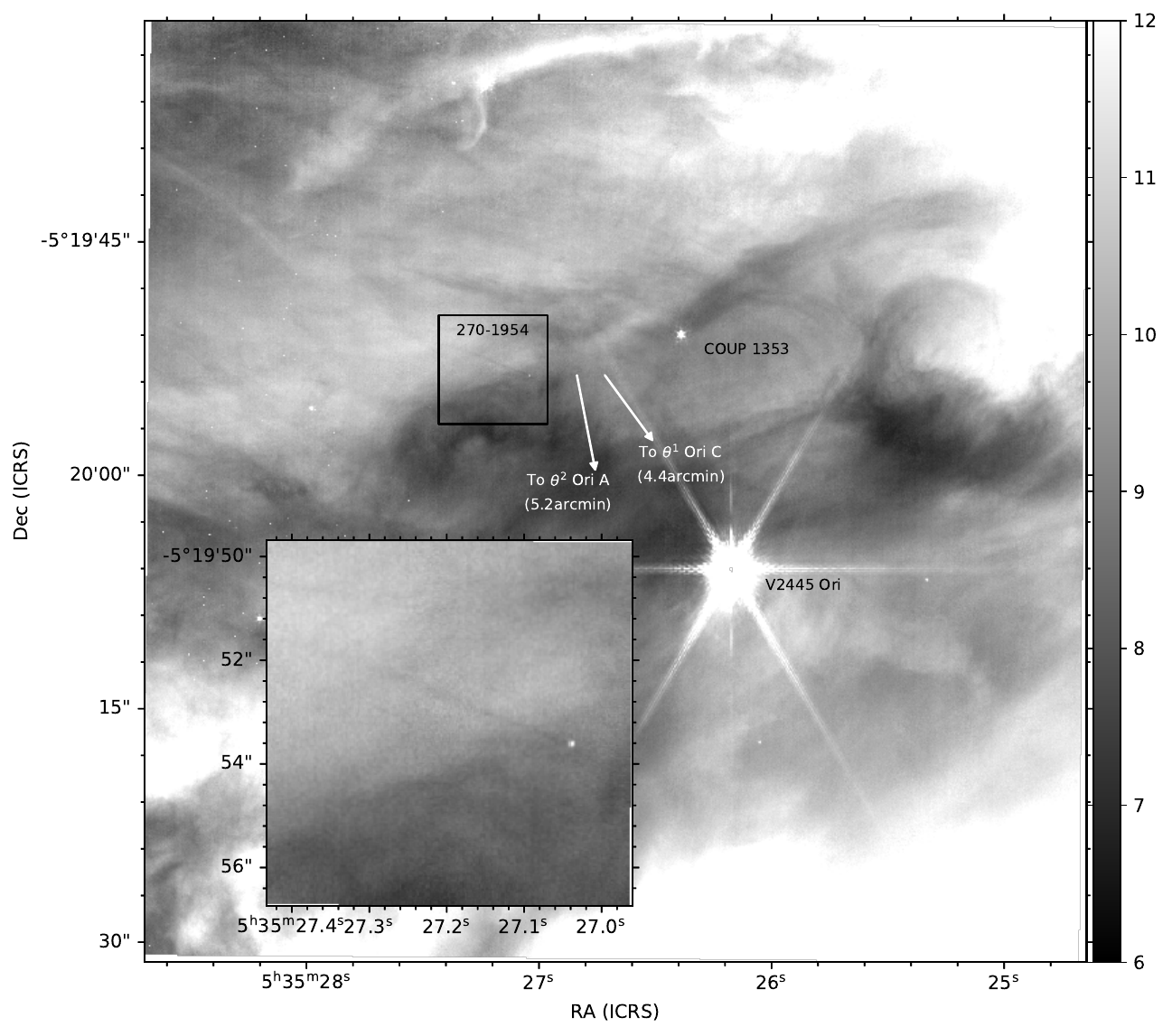}
    \caption{NIRCam F182M observations towards 270-1954, which exhibits an elongated dark tail. The $7\times7\arcsec$ boxed region is shown more clearly in the inset. The arrows and associated labels denote the direction and distance to the main UV sources in the Trapezium Cluster\@. North is up and east left. The intensity scale is in MJy/sr.}
    \label{fig:Overview}
\end{figure*}

In this paper, we use the data of \cite{2023arXiv231003552M} to present the discovery of a highly-extincted brown dwarf candidate that appears to be associated with a a narrow, elongated dark trail. We explore the possibility that the trail might be a physical structure of gas with enhanced maximum dust grain size relative to the ambient medium due to either mass loss from the circum-brown dwarf material, or alternatively due to ongoing Bondi-Hoyle-Lyttleton accretion. If the former, it represents a new instance of radiation-driven mass loss in the outskirts of the Trapezium Cluster, and if the latter it represents a unique instance of ongoing accretion in a high mass star forming region seen to date. Either way this could provide a new view of the impact of environment on a young brown dwarf in a massive star-forming region. 

\section{JWST Observations}
A detailed description of the JWST NIRCam survey of the inner Orion Nebula and Trapezium Cluster that we use is given in \cite{2023arXiv231003552M} and we will not repeat that level of depth here. To summarise, the dataset is a JWST NIRCam \citep{2005SPIE.5904....1R} mosaic spanning 11$\arcmin$ in right ascension and 7.5$\arcmin$ in declination. NIRCam allows simultaneous observations in short- and long-wavelength channels, and for this survey, six pairs of filters were used, namely (in ordering of increasing wavelength): F115W, F140M, F162M, F182M, F187N, F212N (short-wavelength channel), F277W, F300M, F335M, F360M, F444W, and F470N (long wavelength channel). The integration times in all filters was 773.064\,s except for the F115W and F444W pair, which had integration times of 515.365\,s. The spatial resolution in the short-wavelength channel is 31.2775\,mas/pixel, corresponding to approximately 12\,au per pixel at a distance of 390\,pc, and 62.9108\,mas/pixel or $\sim$\,25\,au in the long-wavelength channel. Details of the readout mode set-up and data processing are given in \cite{2023arXiv231003552M}. Unless otherwise specified we will assume 390\,pc as the distance throughout this paper \citep{2022A&A...657A.131M}. 

\section{Introducing 270-1954 and the dark trail}
\label{sec:introductin2701954}

\subsection{Overview}
\label{sec:introoverview}
Figure \ref{fig:Overview} shows our main target in the NIRCam F182M filter (it is also shown in colour composite for illustrative purposes in Figure \ref{fig:colourimage}). It is situated in the outskirts of the Trapezium Cluster, in the Dark Bay region, at approximately, $4.4\arcmin$ (around 0.5\,pc, at 390\,pc) in projected separation to the north-east of the O6V star $\theta^1$\,Ori C and $5.2\arcmin$ from the O9.5IV star $\theta^2$\,Ori A. For reference, the Orion Bar has a closest projected separation from $\theta^1$\,Ori C of more like 0.2\,pc \citep[for papers on the structure of the ONC see e.g.][]{1974PASP...86..616B, 1986ApJ...307..609H, 1995A&A...294..792H, 2009AJ....137..367O, 2017MNRAS.464.4835O, 2017ApJ...837..151O, 2019ApJ...881..130A}  and most proplyds are at even closer distances than that \citep[e.g.][]{2008AJ....136.2136R}

\begin{table}
    \centering
    \begin{tabular}{lcc}
    Object  & RA & Dec \\
    \hline
    270-1954  & 05 35 27.04 & $-$05	19 53.7 \\
    COUP 1353 & 05 35 26.41 & $-$05 19 50.6 \\
    V2445 Ori & 05 35 26.17 & $-$05 20 06.0 \\
    \hline
    \end{tabular}
    \caption{J2000.0 coordinates for 270-1954 and other objects of interest in this work}
    \label{tab:coordinates}
\end{table}

The inner Orion Nebula has an established naming convention for objects that was introduced by \cite{1994ApJ...436..194O} and which is based on J2000.0 coordinates. In this scheme, the name XXX-YYY corresponds to coordinates 05:35:XX.X in right ascension and $-$05:2Y:YY in declination. For objects east of 05:35 and/or north of $-$05:20, an additional digit is introduced at the start of the corresponding part of the label to remove ambiguity. Thus, given the coordinates of the object that is the focus of this paper (Table \ref{tab:coordinates}), we use the identifier 270-1954. 

\begin{figure*}
    \centering
    \includegraphics[width=1.3\columnwidth]{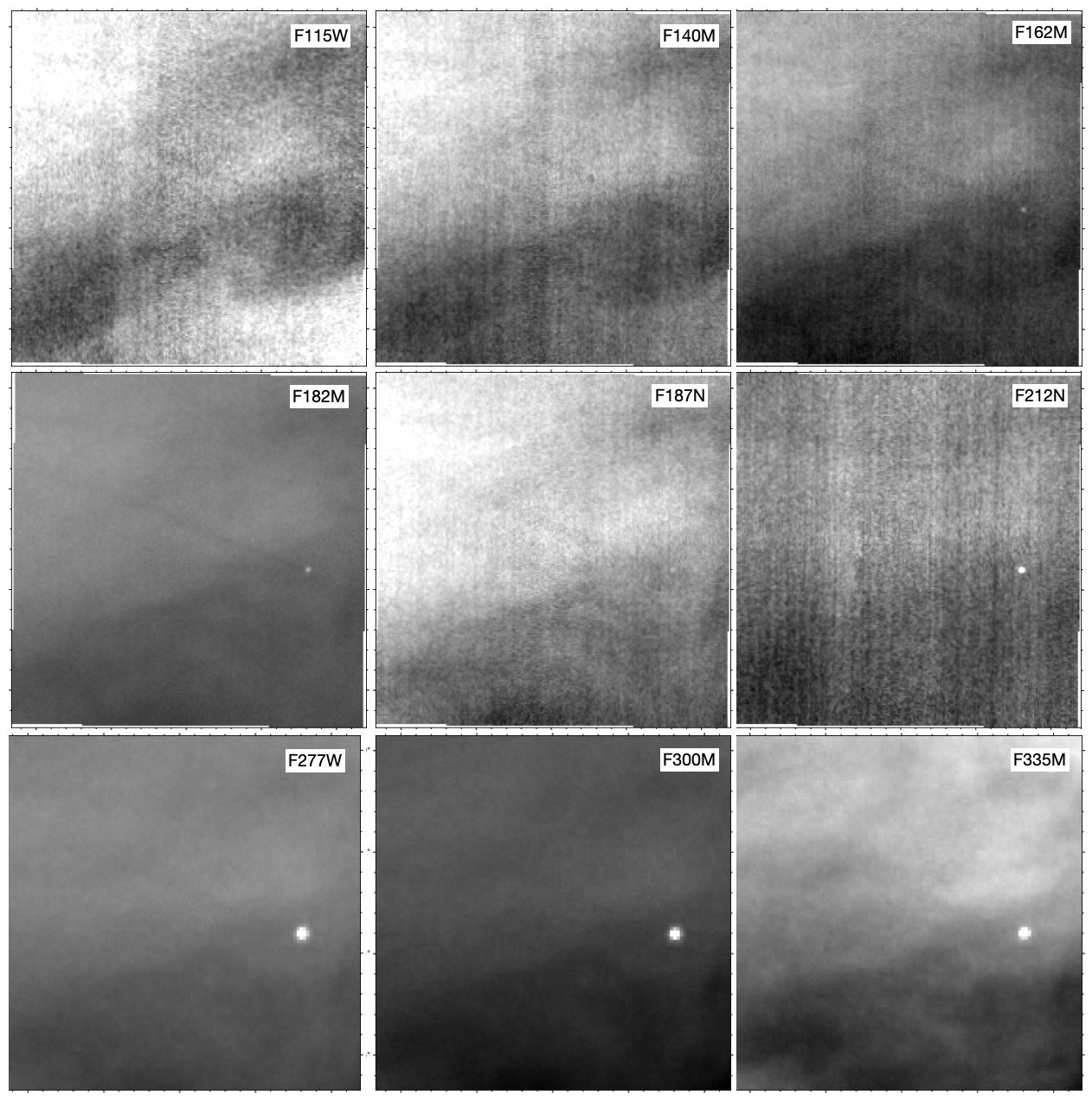}
    \caption{A gallery showing 270-1954 and the associated dark trail in nine JWST NIRCam filters. Each image is $7\times7$ arcsec or $\sim2720\times2720$\,au in size, with North up, East left.}
    \label{fig:Gallery}
\end{figure*}

In Section~\ref{sec:massExt}, we will demonstrate that 270-1954 is likely a highly-extincted, low-mass brown dwarf. The JWST short-wavelength channel images show that it lies at the head of a dark trail extending approximately 4.35\arcsec (1700\,au) in length to its north-east. The trail is about 4-6 pixels ($\sim48-72$\,au) in width, broadening slightly at the end opposite to the point source, making this is a remarkably long and narrow feature. 

270-1954 and the dark trail lie in the so-called Dark Bay region which lies to the north-east and slightly in the foreground of the Trapezium OB stars. The Dark Bay is believed to be a ``fold'' of molecular material which overlays the background PDR and HII{} region, dimming the emission nebulosity and also reddening stars in the Trapezium Cluster seen through the Dark Bay. The optical depth is not extreme, however, as some background galaxies can be seen in the JWST short-wavelength colour composite of \cite{2023arXiv231003552M}.

The nebulosity, both bright and dark, in the region around 270-1954 is complex. There are large-scale dark features extending from the south-east to north-west, with red rims facing towards the Trapezium OB stars and likely illuminated by them. There are also fainter dark wisps and striations roughly perpendicular to these features, which are either relatively broad and variable in width, or with low contrast relative to the surroundings. {We note that the dark feature in the vicinity of COUP 1353 is coincident with a dark ridge observed in the optical with HST (also visible in Figure \ref{fig:colourimage}) so while it is tempting to suggest it is also a physically associated trail we treat it as due to that optical ridge in this paper. }

The dark trail that is the subject of this paper is different, however. The extreme aspect ratio, the straightness, the coincidence of a faint red point source, 270-1954, at the head of the trail and the slight broadening of the trail at the opposite end to the point source (as one would expect for a wake) suggest a possible physical association. It is difficult to quantify the likelihood of chance alignment. We make a very rough initial assessment on the basis that the mosaic is $11\times7$ arcminutes and contains 3000 stars. The probability of one of 3000 randomly distributed stars being within the trail width of either edge of the trail is $\sim0.02\%$ (i.e. 500 trails in the mosaic would give a $10\%$ probability of chance alignment). The stellar density is concentrated closer to the centre of the mosaic, far from 270-1954, which would make the probability of chance alignment even lower. 

\begin{figure}
    \centering
    \includegraphics[width=0.9\columnwidth]{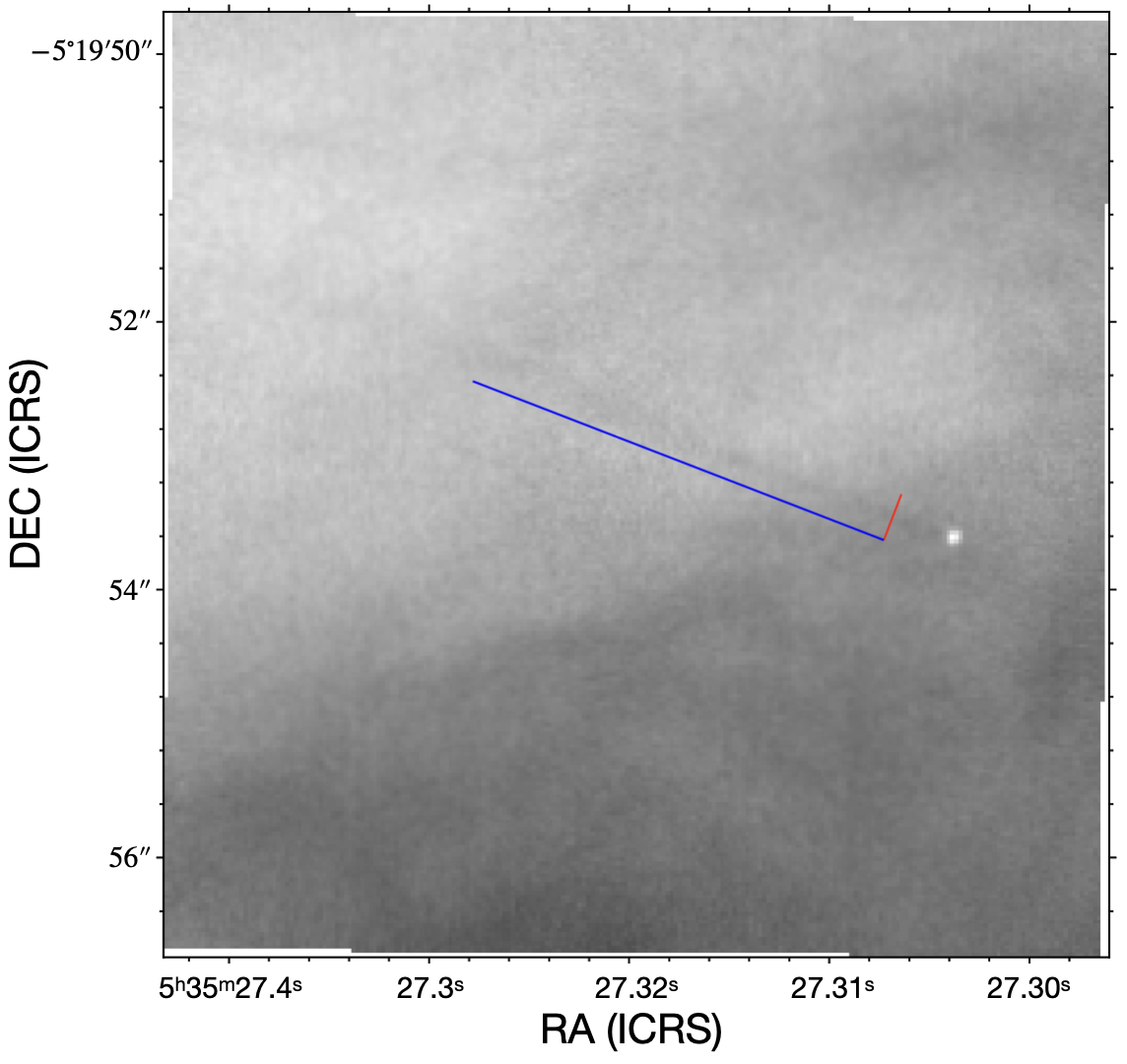}
    \caption{An illustration of the cuts made across the trail associated with 270-1954. The red line is a cut across the trail. From 100 points along the red line we calculate the mean along the blue line parallel to the trail. We plot the resulting profiles in Figure \ref{fig:cuts}.}
    \label{fig:cutlocations}
\end{figure}

A gallery of images of the trail in nine of the JWST NIRCam filters is shown in Figure~\ref{fig:Gallery}. The contrast of the trail against the background nebulosity is there, but noisy, at short wavelengths, with less noise in the F182M. The trail is just visible in F212N and F277W, but is essentially invisible at longer wavelengths.

{Quantifying the contrast between the trail and the ambient nebula is challenging. The ambient nebulosity is quite variable and there is a gradient from north-west to south-east that cuts across the trail. By manual inspection the trail is typically at order the per-cent level fainter than the ambient nebulosity perpendicular to a given point in the trail. However, it is noisy. At some points in a cut across the trail the trail is actually slightly brighter than the immediate ambient nebulosity. }

\begin{figure*}
    \centering
    \includegraphics[width=1.5\columnwidth]{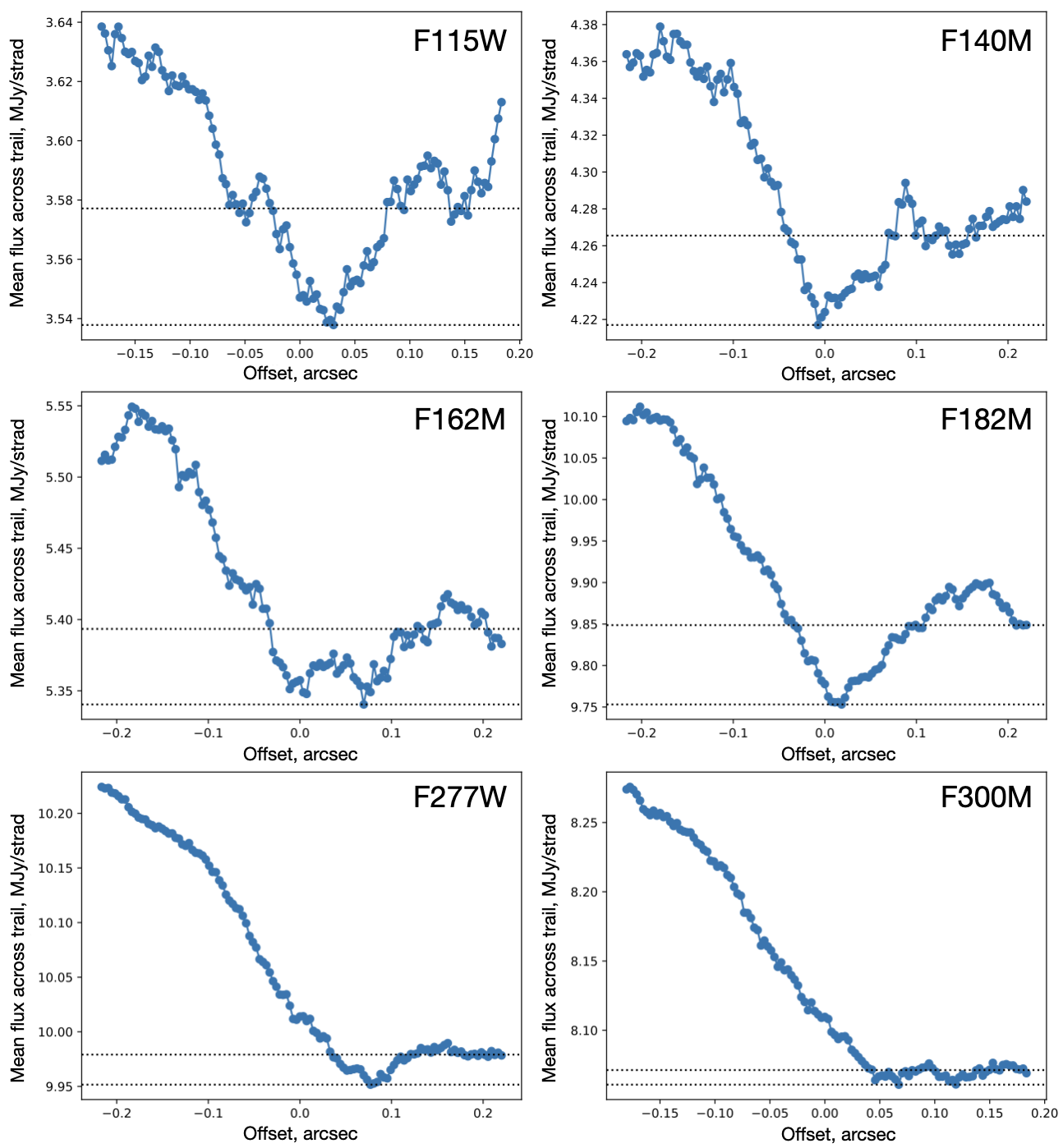}   
    \caption{The profile of the trail perpendicular to it, calculated as a the mean along the length of the trail, for the nine filter images shown in Figure~\ref{fig:Gallery}. We approximate the depth of the trail using the difference between the minimum of the trail and median on the fainter side of the trail, both of which are shown by the dotted lines. Note that zero offset is halfway along the red cut in Figure \ref{fig:cutlocations}. }
    \label{fig:cuts}
\end{figure*}

{We quantify the value of the ``contrast'' by averaging the profile along the trail. Figure~\ref{fig:cutlocations} shows a blue line parallel to the trail and a red line that crosses perpendicular to the trail. We make 200 similar cuts of the red line along the blue. If we average all 200 cuts as a function of position along the red we obtain profiles such as in Figure \ref{fig:cuts}. In those plots an offset of zero is the centre of the red line in Figure~\ref{fig:cutlocations}, negative offset is to the north-west and positive offset to the south-east. Recall that the pixel size is 31.5\,mas and 62.9\,mas in the short and long wavelength filters, but the profiles do not appear blocky at that scale because the cuts are not aligned with the pixel grid axis. There is a gradient in the background nebulosity from north-west to south-east and thus we quantify the depth of the trail using the difference between the faintest part of the trail and the median in the fainter side of the background. That depth is represented by the pair of dotted lines in each panel of Figure \ref{fig:cuts}. This illustrates that the contrast decreases with increasing wavelength as expected from the images.  In order to make an estimate of the uncertainty we employ a bootstrapping approach. We randomly sample 10\,per cent of the 200 cuts 1000 times and calculate the mean and standard deviation of the resulting contrasts (as defined above, using the minimum in the trail and median on the fainter side of the cut). The resulting values are given in Table \ref{tab:depth}.}

\begin{table}
    \centering
    \begin{tabular}{cccccccc}
    \hline
    Filter  & \% drop of min relative  & standard  \\
    & to ambient median & deviation (\%) & \\
    \hline
    F115W     & 1.01 & 0.39 \\
    F140M     & 1.32 & 0.57 \\
    F162M     & 0.79 & 0.32 \\
    F182M     & 0.81 & 0.26 \\ 
    F187N     & 0.8 & 0.34 \\    
    F277W     & 0.55 & 0.71 \\     
    F300M     & 0.73 & 0.55 \\   
    F335M     & 0.43 & 0.31 \\
    \hline
    \end{tabular}
    \caption{The contrast between the tail associated with 270-1954 and the surrounding nebula in various NIRCam filters. The drop in flux in the trail relative to the surroundings is calculated using the difference between the minimum in the trail and the median value in the fainter side of the surrounding nebula (see Figure~\ref{fig:cuts}). {We use a bootstrapping approach to calculate the mean and standard deviation of the contrast from 1000 sets of 20 cuts across the trail.} Note that we exclude the F212N filter here because the profile was substantially noisier (see also Figure \ref{fig:Gallery}).   }
    \label{tab:depth}
\end{table}

\subsection{Brown dwarf mass and extinction estimates from NIRCam photometry}
\label{sec:massExt}
\begin{figure}
    \centering
    \includegraphics[width=\columnwidth]{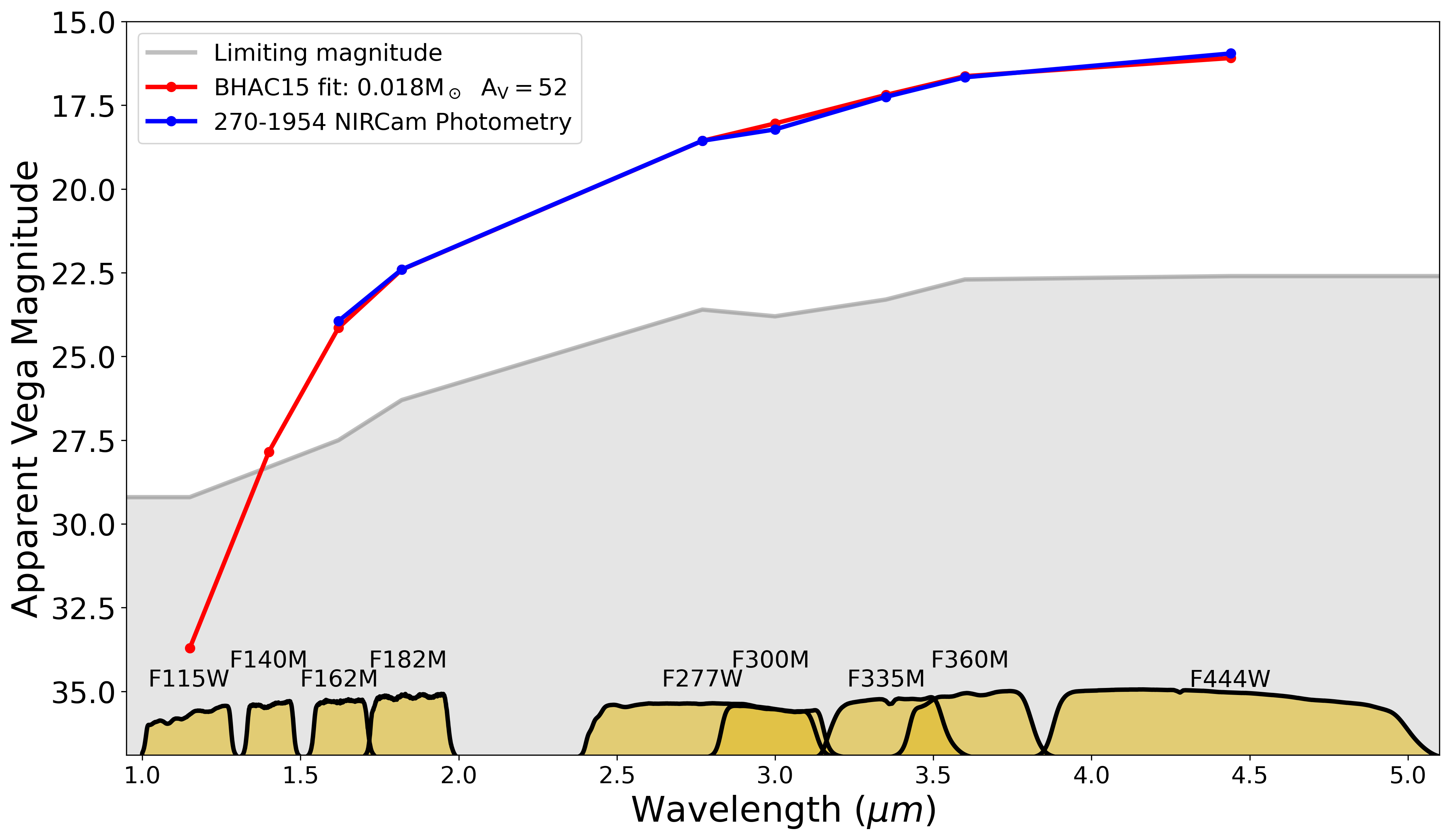}
    \caption{NIRCam SED of 270-1954 (blue line) as sampled by the nine NIRCam medium- and wide-band filters (yellow). The best-fitting BHAC15 model SED \citep{2015A&A...577A..42B} with mass and extinction of 0.018\,$\pm 0.007 M_\odot$ and ${\rm A}_{\rm V} = 52 \pm 9$\,mag is shown in red. The approximate limiting magnitudes of the NIRCam observations in a relatively low nebulosity region are shown in grey \citep{2023arXiv231003552M}.}
    \label{fig:SED}
\end{figure}

To estimate the mass of the 270-1954 point source and the line-of-sight extinction towards it, we use aperture photometry to measure its flux in the seven NIRCam medium- and wide-band filters where it is visible in the JWST images. The blue points and line in Figure~\ref{fig:SED} show the resulting spectral energy distribution. The grey line shows the approximate JWST point source limits determined for relatively low-nebulosity regions of the survey, as appropriate here \citep{2023arXiv231003552M}, while the filter profiles are shown in yellow at the bottom of the plot. 

We fit the SED using the BHAC15 evolutionary models \citep{2015A&A...577A..42B} by varying the mass and extinction, while assuming a nominal age of 1\,Myr, a distance of 390\,pc, and the reddening law from \citep{Wang2019}. The latter was adjusted to use an optical total-to-selective extinction ratio ${\rm R}_{\rm V} = 5.5$, which has been found to be more appropriate for the Trapezium Cluster and ONC compared to the standard value (${\rm R}_{\rm V} = 3.1$) used for the diffuse ISM \citep{Fang2021}. The BHAC15 models were linearly interpolated to cover masses from 0.01\,M$_\odot$ to 1.4\,M$_\odot$ and the extinction was bound between ${\rm A}_{\rm V} = 0$--100. The {\tt scipy.optimize.curve\_fit} package \citep{2020SciPy-NMeth} was used to find the optimum non-linear least squares to fit using the Trust Region Reflective method. 

The best-fitting mass and extinction for 270-1954 are 0.018\,$\pm 0.007 \rm{M}_\odot$ and A$_{\rm V} = 52^{+9}_{-2}$\,mag, respectively. The dominant source of uncertainty in the determined mass stems from the uncertainty in age. We repeated the SED fitting using BHAC15 models for both 0.5\,Myr and 2\,Myr and take the range of mass estimates as the uncertainty for the mass. The upper bound of the uncertainty in the fitted extinction was determined in a similar manner by repeating the SED fitting using the ISM reddening law with ${\rm R}_{\rm V} = 3.1$. The lower bound of the uncertainty was taken as the standard deviation on the fitted extinction, determined by the curve\_fit routine using the Levenberg-Marquardt algorithm.

\begin{figure}
    \centering
    \includegraphics[width=0.9\columnwidth]{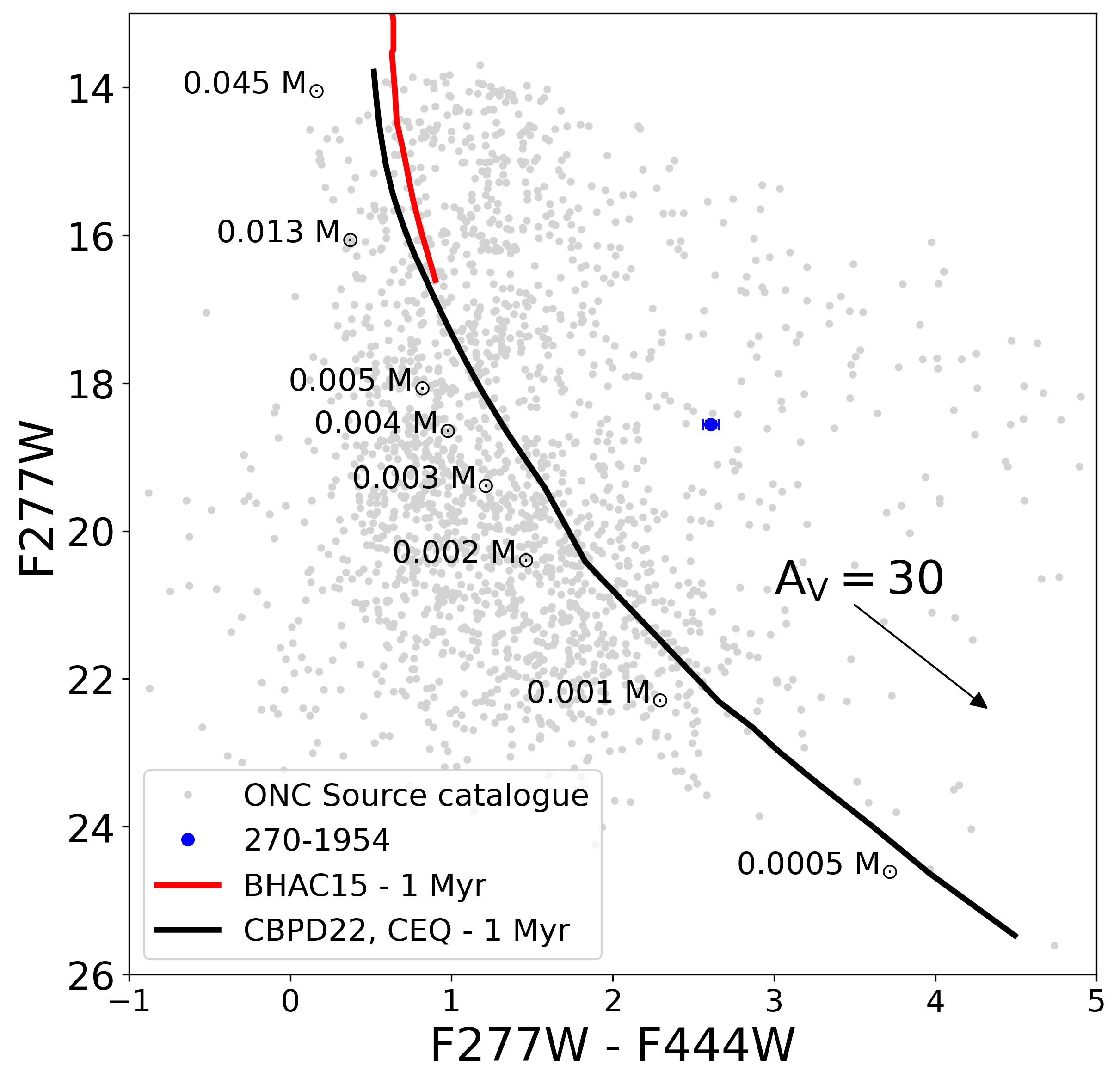}
    \caption{Colour-magnitude diagram of 270-1954 (blue). The full Trapezium Cluster source catalogue is shown in grey \citep{2023arXiv231003552M}. The 1\,Myr isochrone for the BHAC15 evolutionary models is shown in red \citep{2015A&A...577A..42B}, while the equivalent 1\,Myr isochrone for the CBPD22~CEQ evolutionary models \citep{Chabrier2023}, which combine the atmospheric models from ATMO 2020 \citep{Philips2020} with the a new equation-of-state for dense hydrogen-helium mixtures \citep{Chabrier2021}, is shown in black. An approximate unreddened mass scale is shown.}
    \label{fig:CMD}
\end{figure}

Figure~\ref{fig:CMD} shows a F277W vs F277W\,$-$\,F444W colour magnitude diagram for Trapezium Cluster sources as grey points and 270-1954 in blue. A 1\,Myr isochrone for the BHAC15 \citep{2015A&A...577A..42B} evolutionary models for brown dwarfs and stars is shown in red, while the CBPD22~CEQ \citep{Chabrier2023} evolutionary models for brown dwarfs and planetary-mass objects at 1\,Myr are shown in black. An ${\rm A}_{\rm V}=30$ reddening vector and an approximate unreddened mass scale are over-plotted. It can be seen that if 270-1954 is dereddened parallel to the reddening vector, it will end up in the low-mass brown dwarf mass range. 

\begin{figure*}
    \centering
    \includegraphics[width=1.6\columnwidth]{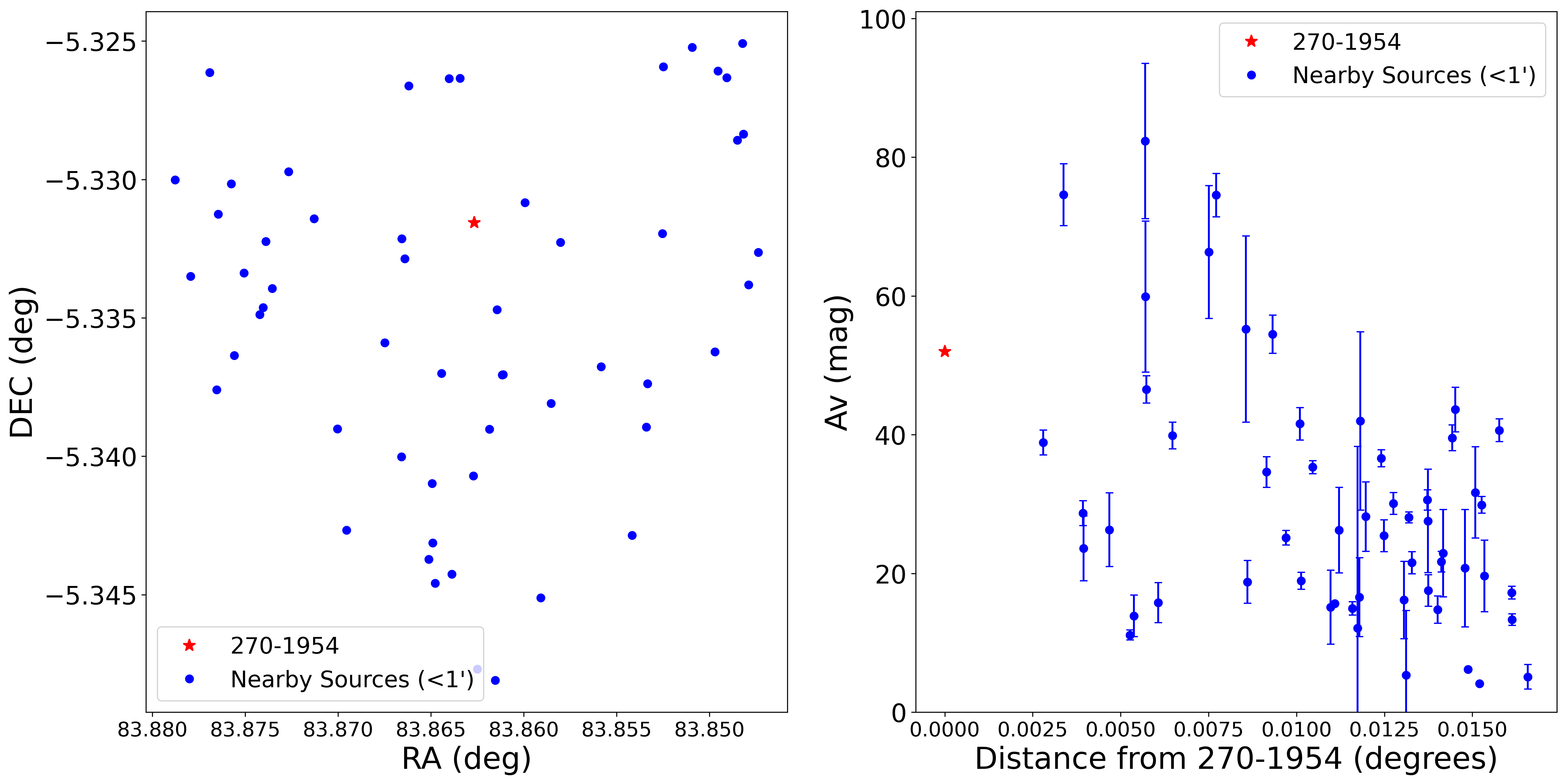}
    \caption{The left-hand panel shows the locations of the 55 point sources within one arcminute of 270-1954, truncated on the northern side due to the declination limit of the survey. The right-hand panel plots the A$_{\rm V}$ for each source as determined by SED fitting as a function of its projected separation from 270-1954. }
    \label{fig:nearbyAVs}
\end{figure*}

Figure \ref{fig:nearbyAVs} shows the ${\rm A}_{\rm V}$s of 55 sources within one arcminute of 270-1954 (the sample is truncated at the northern edge due to the declination limit of the survey). The extinctions for these sources were determined using the same SED fitting method described above. Although the foreground extinction towards the Trapezium Cluster and ONC as a whole is typically low (A$_{\rm V} \sim 1$\,mag) \citep{Scandariato2011}, it can be seen from Figure~\ref{fig:nearbyAVs} that 270-1954 and surrounding sources are located in a region of higher extinction associated with the Dark Bay, as seen in the NIRCam images.

\begin{figure*}
    \centering
    \includegraphics[width=0.7\linewidth]{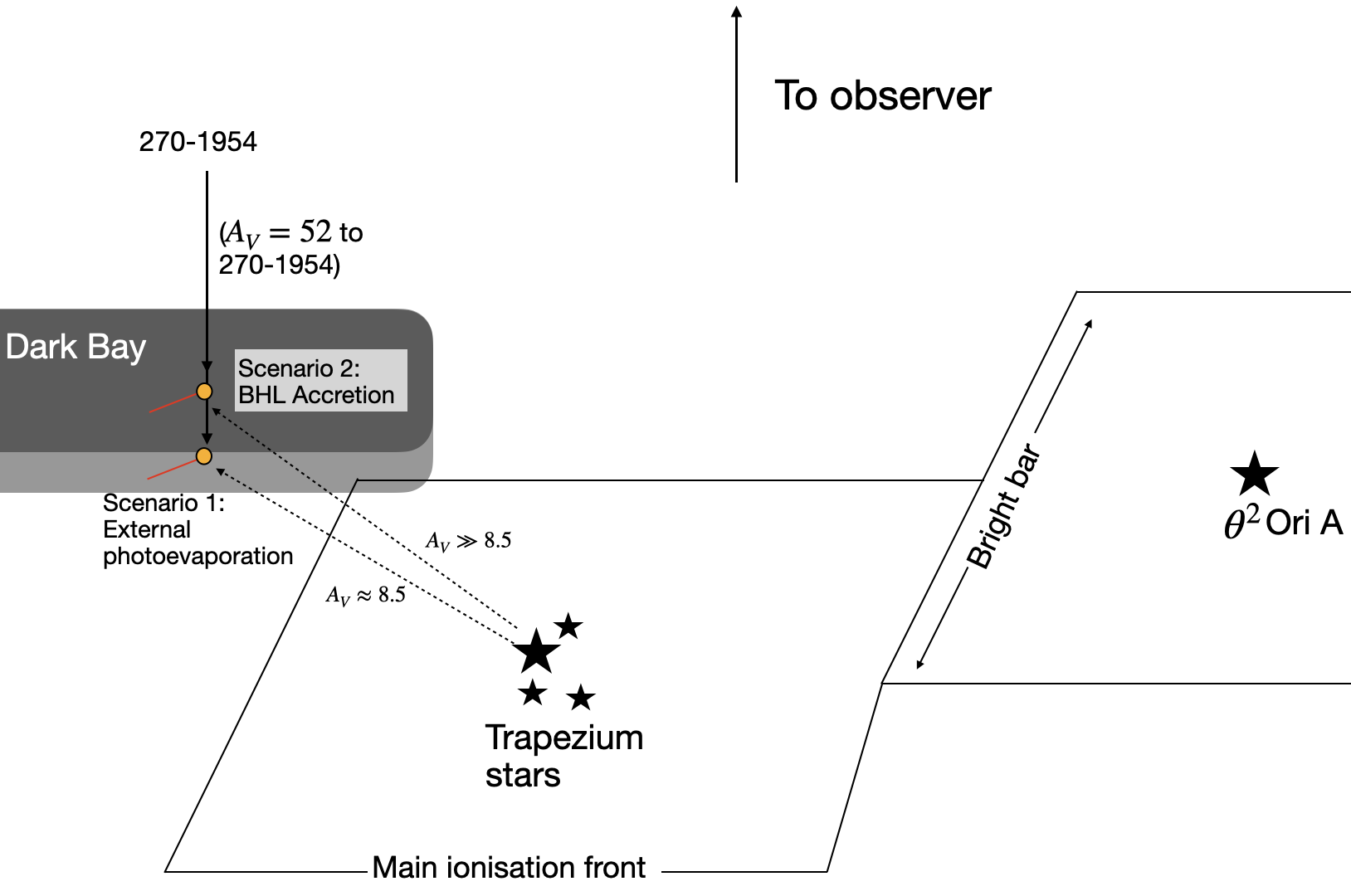}
    \caption{{A cartoon illustrating the possible geometry of the region and 270-1954. The trapezium stars sit within a bowl-like region, with the main ionization front of the nebula on the far side from the observer. The Orion Bright Bar corresponds to where the main ionization front curves along the line of sight. The Dark Bay against which 270-1954 is coincident lies slightly in the foreground. The 270-1954 brown dwarf is denoted by the yellow circle and the trail by the extended red line. There are two brown dwarf/trail cartoons on the figure representing two possible scenarios.  We will demonstrate that  270-1954 could be leaving a trail due to a weak external photoevaporative wind even if the extinction from 270-1954 to the trapezium stars is around $A_V=8.5$ and the FUV field is 1\,G$_0$, illustrated by the lighter shaded part of the Dark Bay in the cartoon and  explored in detail in section \ref{sec:winddriving}. Alternatively in section \ref{sec:BHL} we will also discuss the possibility that it may be  a Bondi-Hoyle-Lyttleton accretion wake if it were substantially more shielded from UV irradiation by the Dark Bay. }}
    \label{fig:overviewCartoon}
\end{figure*}

\subsection{Interpreting the trail}
Unlike the strongly-irradiated proplyds, which have a cusp-tail ionisation front morphology that points towards the UV source illuminating them, the trail of 270-1954 is {not directed} towards $\theta^1$\,Ori C nor $\theta^2$\,Ori A. Furthermore, it makes an angle of around 25.5 degrees from the radial line joining 270-1954 to the nearest bright star on the plane of the sky, the K6 variable star V2445\,Ori (see Figure \ref{fig:Overview}) ruling out the casting of a shadow by that source. This immediately leaves us with only two possibilities: i) the apparent coincidence between 270-1954 and the very tip of an exceptionally isolated, long, straight, and narrow wisp of darker nebulosity is simply that, a line-of-sight chance alignment {(suggested to be unlikely in \ref{sec:introoverview})} or ii) it is a trail of material left in the wake of 270-1954 as it propagates through the interstellar medium (ISM). The main goal of the rest of this paper is to determine the plausibility of the latter scenario. 

{In the rest of this paper we will demonstrate that a trail of dust slightly larger than that in the ambient medium  could explain the trail, even at high visual extinction associated with the Dark Bay. We will further demonstrate that the two most plausible scenarios for producing the trail are dust entrained by external photoevaporation {or gas/dust left in a} Bondi-Hoyle-Lyttleton accretion wake. A summary of the approximate geometry of the region and the location of 270-1954 in those two scenarios is given in Figure \ref{fig:overviewCartoon}, which we will now proceed to explore in detail.}

\section{Demonstrating a highly extincted dark trail is observable with simple radiative transfer models}
\label{sec:basicRT}
In \ref{sec:massExt}, we inferred a high extinction of A$_{\rm V}\sim 52$ towards the 270-1954 point source. The immediate challenge for interpreting the trail as a physical entity associated with 270-1954 rather than chance alignment with a dark streak of nebulosity is to ask how a trail can be visible if it too is similarly extincted. {In the first instance we want to demonstrate whether this is at all plausible before concerning ourselves with the possible physical mechanisms that might be responsible for the trail. }

\begin{figure}
    \centering
    \includegraphics[width=\columnwidth]{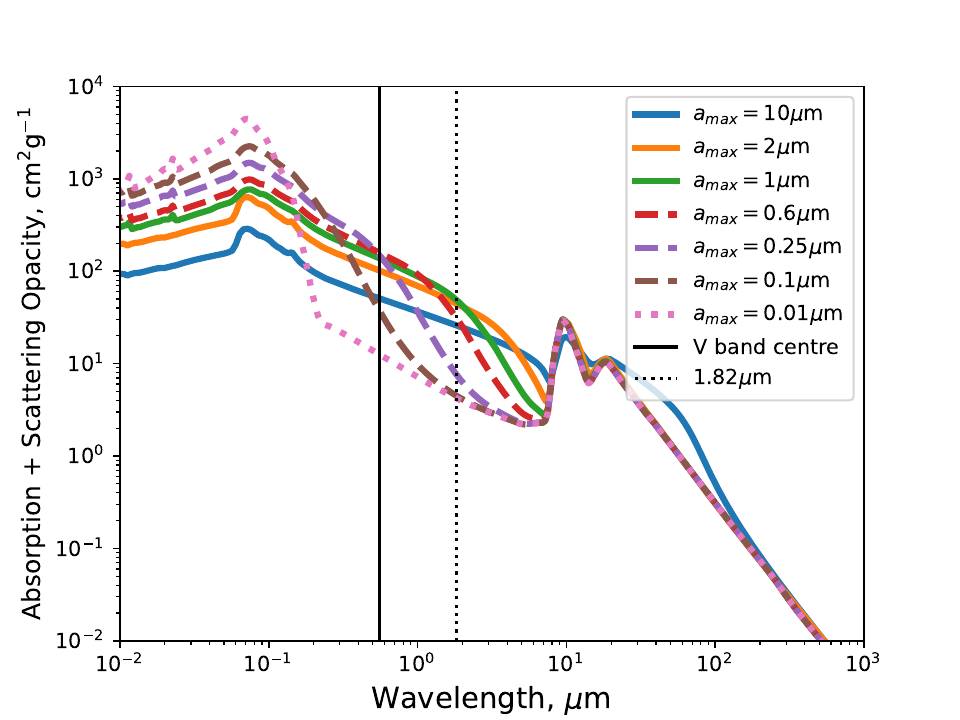}
    \includegraphics[width=\columnwidth]{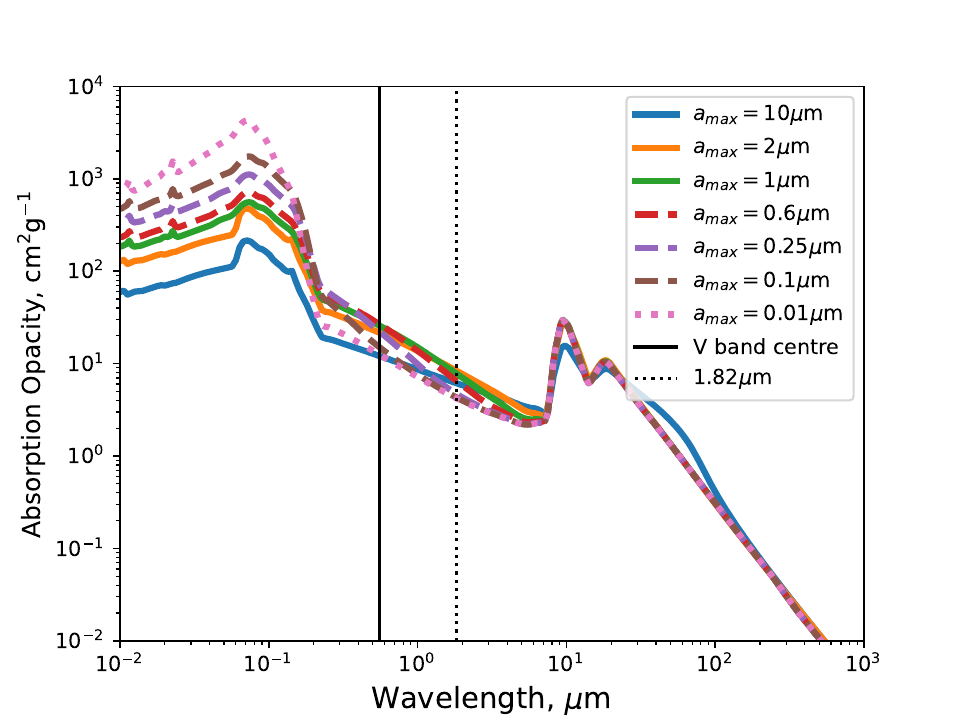}  
    \includegraphics[width=\columnwidth]{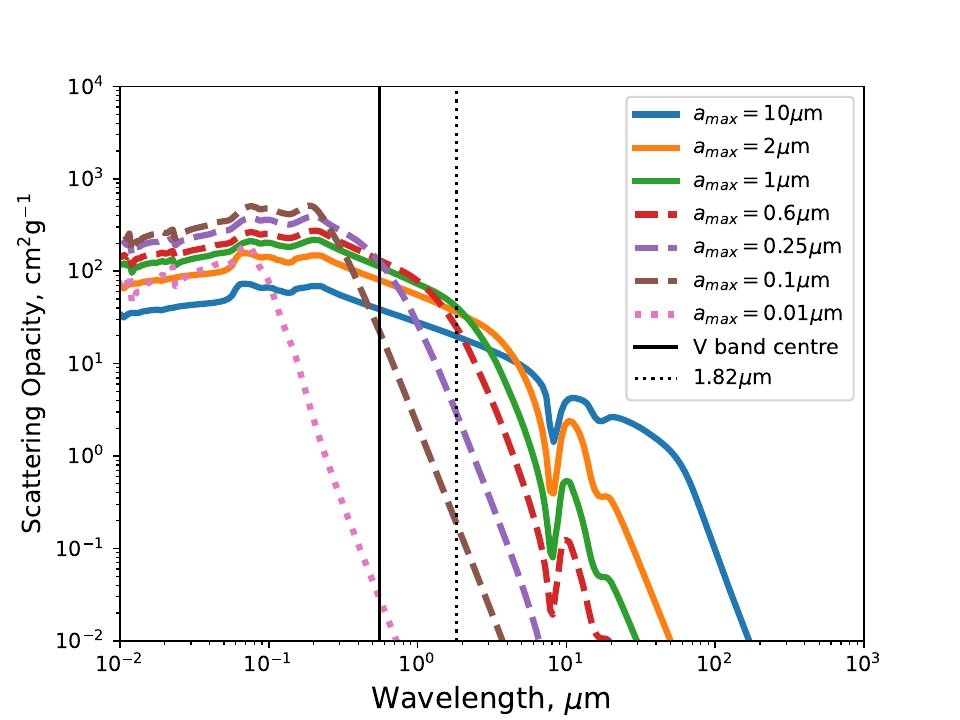}       
    \caption{The total (absorption + scattering, upper panel),  absorption-only (middle panel) and scattering-only (lower panel) opacities as a function of wavelength for \protect\cite{1984ApJ...285...89D} silicates with a $q=3.5$ power law size distribution and maximum grain sizes given in the legend. A slight enhancement in the maximum grain size up to at most around $\sim2$\,$\mu$m leads to a significant increase in the scattering opacity in the near-infrared while always maintaining high extinction in the visible. These curves were computed using utilities in the \textsc{torus} radiative transfer code \protect\citep{2019A&C....27...63H}. }
    \label{fig:opacities}
\end{figure}

At near-infrared ($\sim$\,micron) wavelengths and low temperatures, thermal emission from dust is negligible, so the light we see in the region is either reflected by dust grains in the cloud, or is background emission passing through the cloud. Scattering is potentially very important at these wavelengths, although that depends sensitively upon the maximum grain size in the cloud. Figure~\ref{fig:opacities} shows the total, absorption-only, and scattering-only opacities per gram of cloud material (gas and dust) for \cite{1984ApJ...285...89D} silicate distributions with different maximum grain sizes and a standard ISM power-law size distribution with $q=3.5$. The scattering opacity is very sensitive to the maximum grain size at micron wavelengths. The maximum grain size in the ISM is usually sub-micron \citep{1977ApJ...217..425M}, so if 207-1954 were trailing larger micron-sized grains (we discuss possible mechanisms for doing so below), then the scattering opacity at $\sim$micron wavelengths would rise by over an order of magnitude. Note that if the grains get sufficiently large ($\sim10\,\mu$m) then the scattering opacity distribution flattens and starts reducing again. 

If the dominant source of light is background emission from the \HII{} region and PDR passing through the cloud, this might therefore enable an extincted trail with an enhancement in its maximum grain size up to $\sim$ micron sizes to appear slightly darker relative to the surroundings, by scattering light along the line-of-sight ``out of the trail'' -- recall from \ref{sec:introductin2701954} we only require the trail to be around 1~per cent fainter. We now explore this concept in a simple 1D radiative transfer model. 

\subsection{Radiative transfer model details}
\label{sec:RTmodel}
\begin{figure}
    \centering
    \includegraphics[width=\columnwidth]{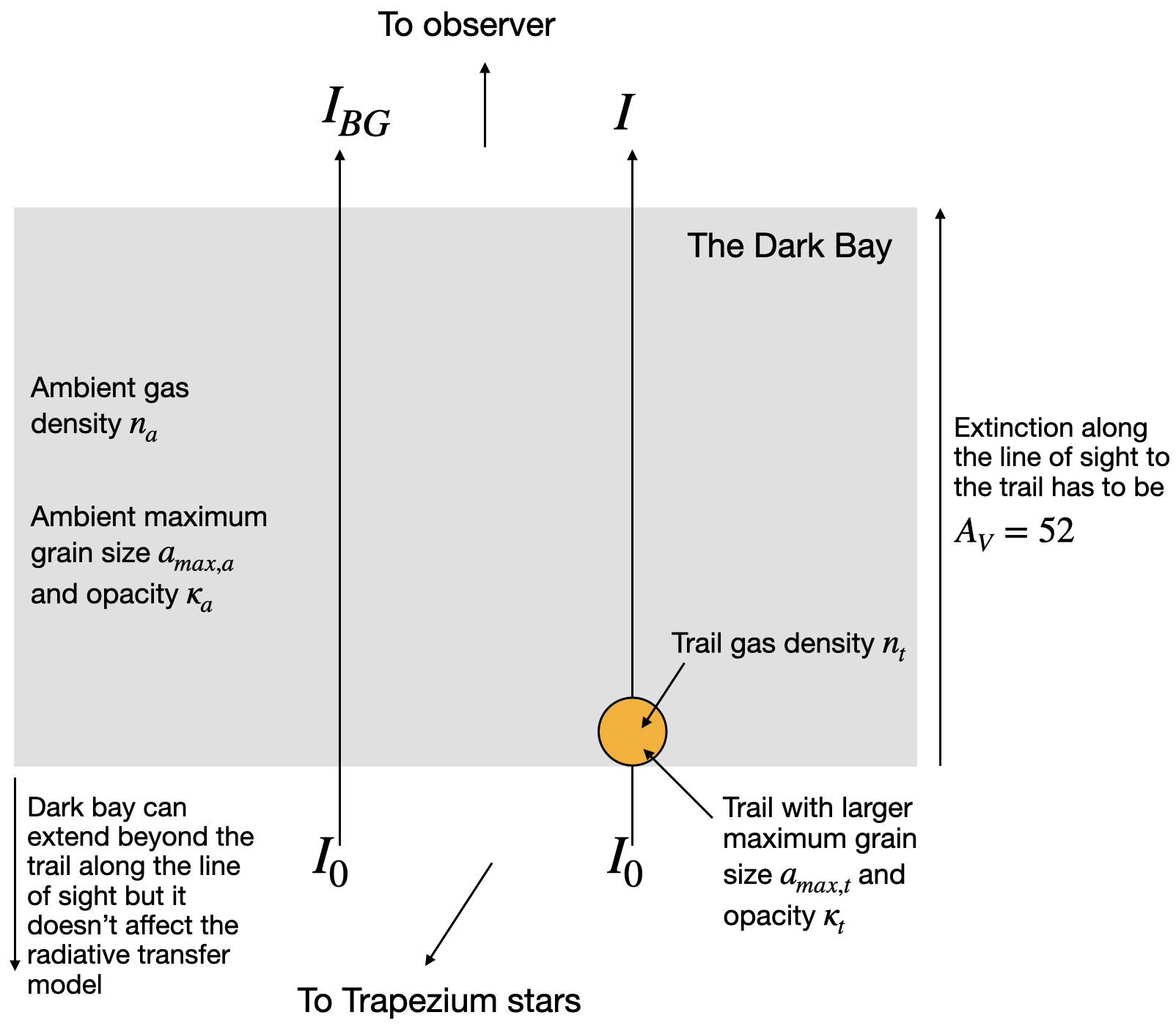}
    \caption{{A schematic of the radiative transfer model we consider. The box is illuminated from below with a uniform intensity $I_0$. A line-of-sight through the ambient medium has an observed intensity $I_{BG}$. A line-of-sight through the trail includes a segment with a different maximum grain size representing the trail and an observed intensity $I$. Note that although the radiative transfer model places the trail at the far edge of an $A_V=52$ slab, the Dark Bay can extend further in reality, it simply changes the value of $I_0$, which is arbitrary in our model since we consider the contrast between the trail and ambient medium. Also note that although the trail andambient densities are given different symbols here, in most of our calculations the medium gas density is considered uniform ($n_a=n_t$). }}
    \label{fig:RS_schematic}
\end{figure}

Firstly, our radiative transfer model has to satisfy the empirical inference that the extinction is A$_{\rm V} = 52$, and this has to hold regardless of our choice of cloud density. The extinction is approximately the optical depth
\begin{equation}
    {\rm A}_{\rm V} \approx 1.085\tau =  1.085 \kappa_a \rho_a L
\end{equation}
where $\kappa_a$ is the opacity per gram of gas in the ambient Dark Bay (i.e. not in the trail), $\rho_a$ the gas density in the ambient Dark Bay, and $L$ the path length through the medium (i.e. to the trail), see Figure \ref{fig:RS_schematic} for a schematic.

Assuming a V-band opacity of around 100\,cm$^2$\,g$^{-1}$ and an extinction of around A$_{\rm V} = 52$\,mag, we relate the approximate required path length $L$ of 207-1954 in the cloud to the cloud density with the scaling
\begin{equation}
    L = 0.46\,\textrm{pc}\left(\frac{n_a}{10^5\,\textrm{cm}^{-3}}\right)^{-1}. 
    \label{equn:Ldens}
\end{equation}
{This is the length of constant density material required in the foreground of 207-1954 to give the observed extinction. We then consider two paths through this medium, as illustrated in Figure \ref{fig:RS_schematic}. One path is purely through the constant density medium (with intensity $I_{BG}$ in Figure \ref{fig:RS_schematic}), the other path (with intensity $I$ in Figure \ref{fig:RS_schematic}) includes a small component representing the trail with the same constant density but larger maximum grain size and hence higher scattering opacity drawn from the models in Figure~\ref{fig:opacities}.}






Along both paths through the cloud, we solve the radiative transfer equation for the change in specific intensity $I_\nu$ along some path length $\textrm{d}l$
\begin{equation}
    \frac{\textrm{d}I_\nu}{\textrm{d}l } = j_\nu - \kappa_{\textrm{tot}}\rho I_\nu
\end{equation}
where $j_\nu = B_\nu(T) \kappa_{\textrm{abs}}\rho$ is the emission coefficient, $B_\nu(T)$ is the Planck function at temperature $T$, $\kappa_{\textrm{abs}}$ is the absorption opacity, and $\kappa_{\textrm{tot}}$ is the total absorption plus scattering opacity. When the ambient maximum grain size is sufficiently small (less than about 0.3\,$\mu$m) we can neglect the increase in intensity due to scattering from the surrounding cloud into the line of sight that intersects the trail because scattering in the near-infrared will be weak (Figure~\ref{fig:opacities}). We assume that scattering in the trail, which is very narrow, is not sufficient to significantly alter the intensity in the ambient part of the cloud. Finally, we also assume that forward scattering from the trail material into the line-of-sight of the trail is negligible given that the angular size is so small (see Appendix~\ref{sec:phasefunction}). Overall then, we are making the simplifying assumption that scattering only acts as a basic opacity source along the line-of-sight and that the further propagation of that scattered light is assumed to have a negligible effect on the surroundings. 

This is a simple model, but our objectives are not necessarily to try and infer the true parameters of the system, only to determine whether an extincted trail would be observable in the near-infrared. For our main calculations, we assume that the maximum grain size in the ambient ISM is 0.25\,$\mu$m and power law of the size distribution is $q=3.5$ \citep{1977ApJ...217..425M}, though we explore the effect of different ambient grain distribution in \ref{sec:RTmodels}.  

\begin{figure}
    \centering
    \includegraphics[width=\columnwidth]{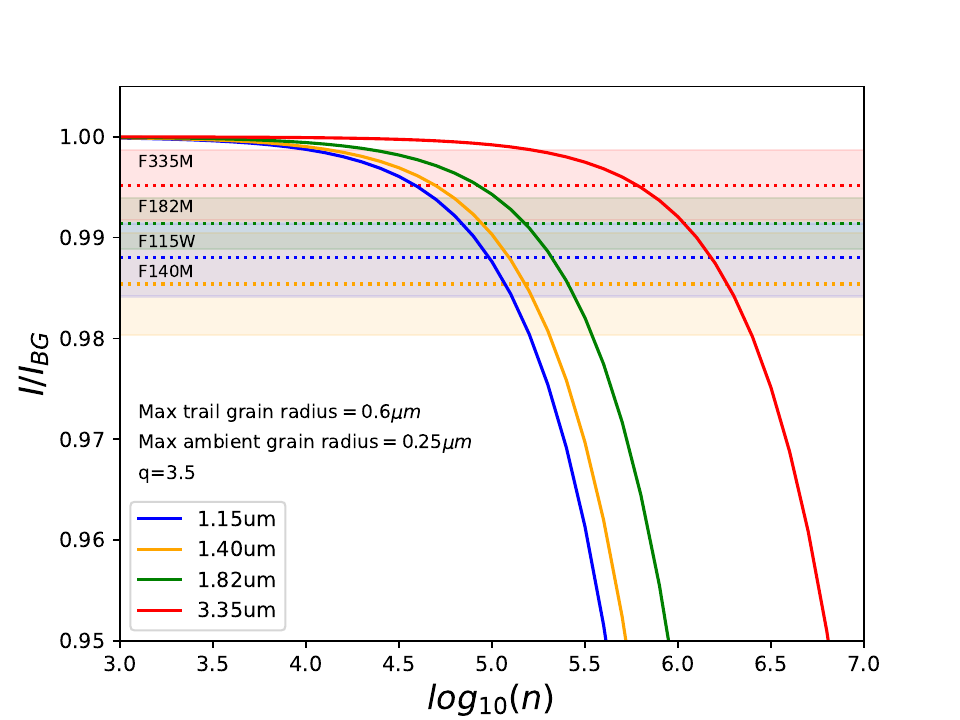}
    \vspace{-0.4cm}
    
    \includegraphics[width=\columnwidth]{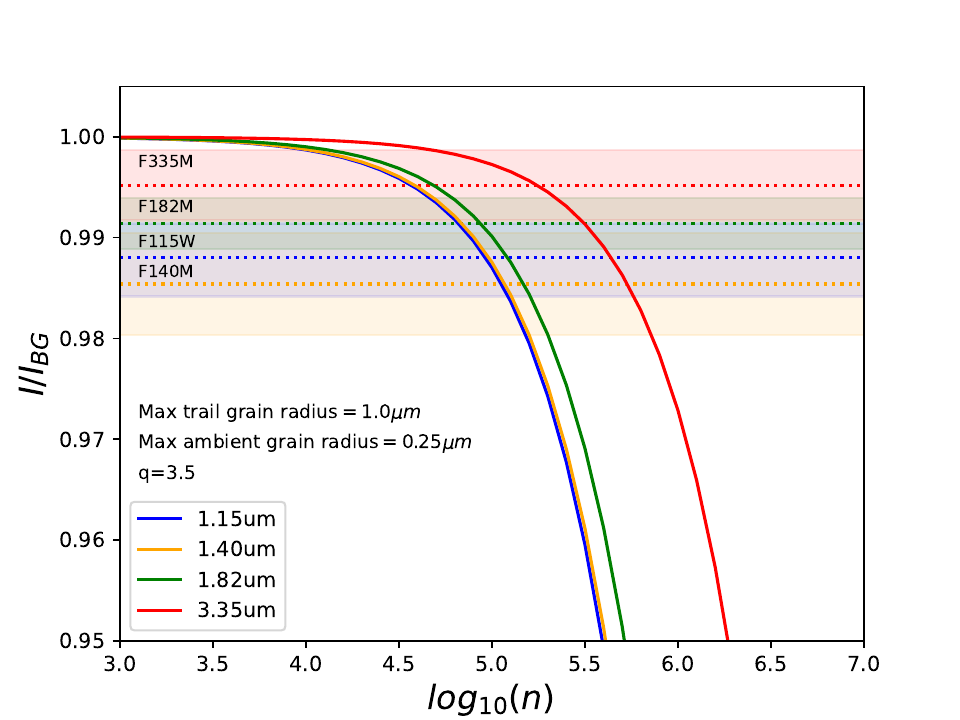}    
    \vspace{-0.4cm}
    
    \includegraphics[width=\columnwidth]{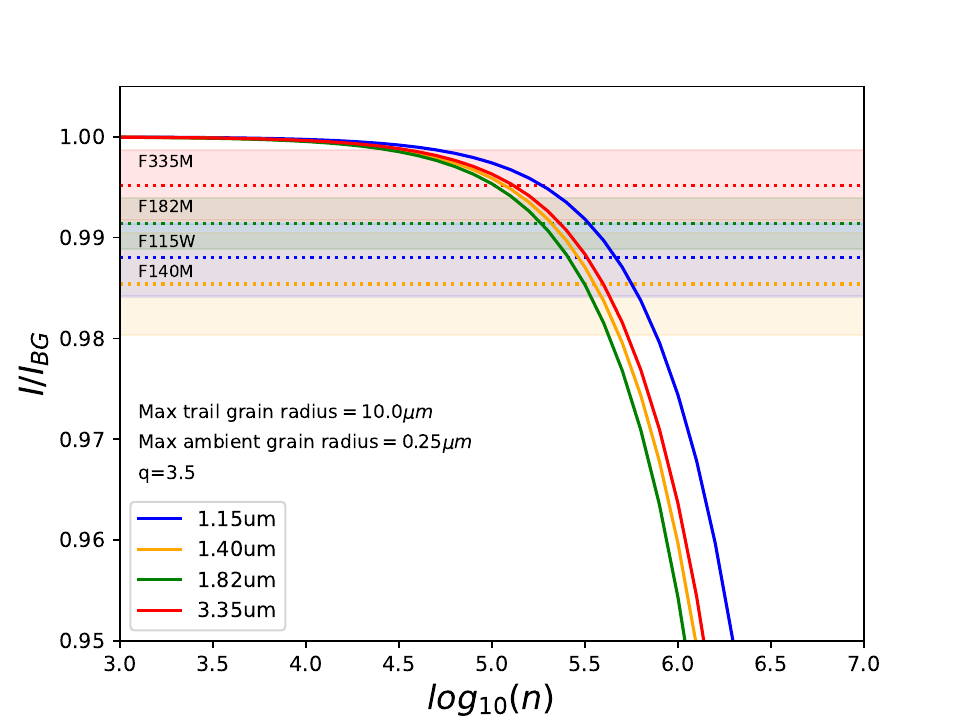}      
    \caption{The ratio of the trail-to-ambient intensities in our basic radiative transfer model as a function of the cloud number density. The panels are for the maximum grain size in the trail of 0.6, 1, and 10\,$\mu$m from top to bottom. Each has a power law of the grain size distribution $q=3.5$. The different coloured lines denote the central wavelengths of different JWST filters. The horizontal dotted lines show the approximate observed contrasts in the F115W, F140M, F182M, and F300M with the shaded areas representing $\pm$1 standard deviation. The upper two panels can reproduce the relative contrast between a LOS along the trail and ambient medium in the different filters for densities $\sim10^5$\,cm$^{-3}$. The lower panel cannot within the 1\,$\sigma$ errors, incorrectly predicting that the F115W filter would have the weakest contrast.  }
    \label{fig:Iratios}
\end{figure}

\subsection{Initial expectations as to the plausible properties of the Dark Bay}
\label{sec:expectations}
{Before coming to the results of the radiative transfer model, we note that some scenarios in the framework set up in \ref{sec:RTmodel} can immediately be expected to be unlikely. The entire Huygens region of the inner Orion Nebula is around 1\,pc across and typically at much lower extinction than the Dark Bay \citep[e.g.][]{2015A&A...582A.114W}, which is $\sim0.25$\,pc across. For low uniform densities the Dark Bay would have to be incredibly extended to give the required extinction, e.g. 46\,pc for $n_a=10^3$\,cm$^{-3}$ with an associated aspect ratio of the Dark Bay itself approaching 200, whereas the more extreme observed filaments have aspect ratio $\sim 50$ \citep{2024A&A...686L..11W}. Although the reality will be that the cloud is not a uniform medium, this argument does imply that the majority of the line of sight extinction cannot be due to low density gas in the Dark Bay. 


{Furthermore, if we consider the Jeans' length of the cloud, it is smaller than the cloud extent from equation \ref{equn:Ldens} for 10\,K gas at densities $<10^7$\,cm$^{-3}$, with the difference between the Jeans' length and cloud length increasing in magnitude at lower densities owing to the linear scaling of the cloud length with density and $n_a^{-1/2}$ scaling of the Jeans' length. The Dark Bay is star forming, and Jeans' analysis is very simple compared to the reality of a turbulent molecular cloud, but provides another suggestion that densities in the Dark Bay responsible for most of the extinction are likely to be high, at the level of $10^5-10^7$\,cm$^{-3}$. }

{We finally note that if the Dark Bay were unstable and a circular sheet, the free fall time   dependency on the aspect ratio (i.e. length) and density cancel \citep[following][]{2012ApJ...744..190T} giving a constant free-fall time as a function of density of $\sim0.1$\,Myr. }

\subsection{Radiative transfer model results}
\label{sec:RTmodels}
Figure~\ref{fig:Iratios} shows the intensity ratio between the line of sight traversing the trail, and line of sight through the ambient medium (as illustrated in Figure \ref{fig:RS_schematic}) as a function of the density of the medium. We include three panels with different maximum grain sizes in the trail of 0.6, 1 and 10\,$\mu$m. The horizontal dotted lines in each panel show the approximate flux drop in the F115W, F140M, F182M and F300M filters, with the shaded areas representing $\pm1$ standard deviation (see Table \ref{tab:depth}).  A good model would see the four solid lines cross the dotted lines/shaded regions at a consistent density. From Figure~\ref{fig:Iratios}, for a maximum grain size in the trail of $0.6\,\mu$m and $1\,\mu$m the models roughly reproduce the observed trail/ambient contrast for a medium density of just over $10^5$\,cm$^{-3}$, which is broadly consistent with the arguments made in \ref{sec:expectations} that the density should be high. For a maximum grain size in the trail of $10\,\mu$m the agreement breaks down due to the flattening of the opacity profile (Figure~\ref{fig:opacities}) with the model wrongly predicting that  in the F115W filter the trail would have the lowest contrast with the surroundings.

We also explored introducing a gas density enhancement into the trail. For example, Figure~\ref{fig:Iratio_rhoEnhance} shows that with a density enhancement of a factor 10, the best  observed contrast between the trail and surroundings is retrieved for an ambient density of $<10^4$\,cm$^{-3}$. However as discussed in \ref{sec:expectations} such a density is approaching the regime where the Dark Bay becomes implausibly extended to provide the high visual extinction. So not only can a uniform medium with only a maximum dust grain size enhancement reproduce the contrast between trail and ambient medium (Figure~\ref{fig:Iratios}) our models suggest that there should not be a large gas density enhancement in the trail.

\begin{figure}
    \centering
    \includegraphics[width=\columnwidth]{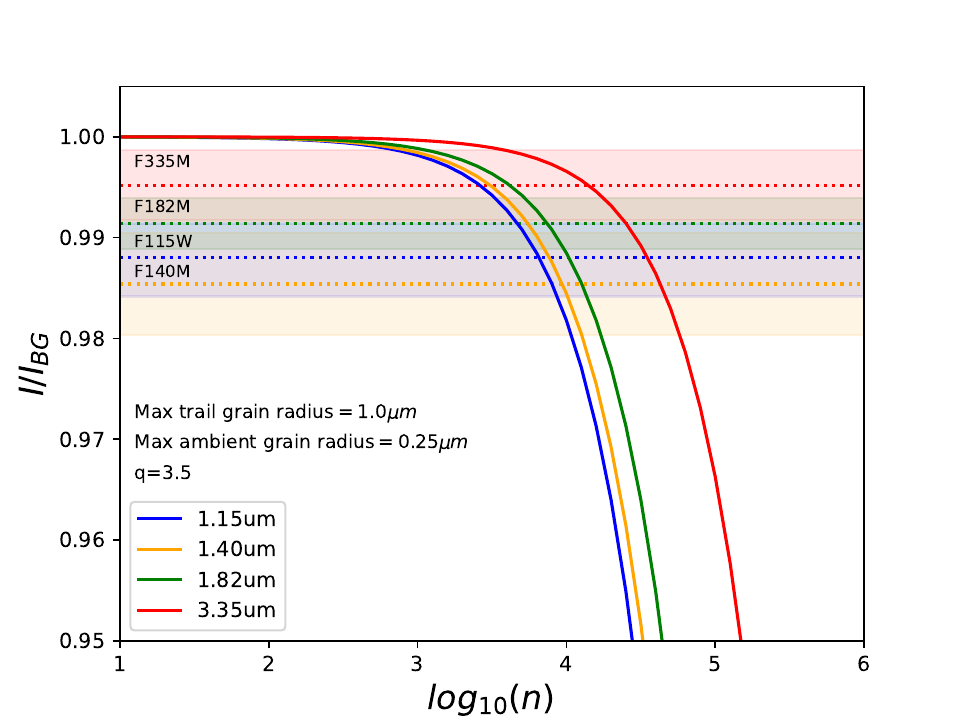}
    \caption{The ratio of the trail-to-ambient intensities as a function of the cloud number density, similar to Figure \ref{fig:Iratios}. In this case, the trail also has a density enhancement of a factor $10\times$ the ambient density and the maximum grain size in the trail is 1\,$\mu$m. So a gas density enhancement in the trail pushes down the required ambient cloud density to give the observed contrast. }
    \label{fig:Iratio_rhoEnhance}
\end{figure}

A caveat to the radiative transfer model so far is that we considered the case of a maximum grain size enhancement in the trail relative to fairly ``standard'' ISM dust in the surrounding medium. That is, a power law size distribution $q=3.5$ and $a_{\textrm{max}}=0.25\mu$m which corresponds to a reddening ${\rm R}_{\rm V} \sim 3.3$. \cite{Fang2021} presented a spectroscopic survey of stars in the Trapezium cluster, comparing their analysis with ${\rm R}_{\rm V} = 5.5$ and ${\rm R}_{\rm V} = 3.1$ and finding that the former performed better. The reddening may therefore be higher in general in the region (indeed we adopted ${\rm R}_{\rm V} = 5.5$ in our analysis of section \ref{sec:massExt}).

\begin{figure}
    \centering
    \includegraphics[width=0.99\columnwidth]{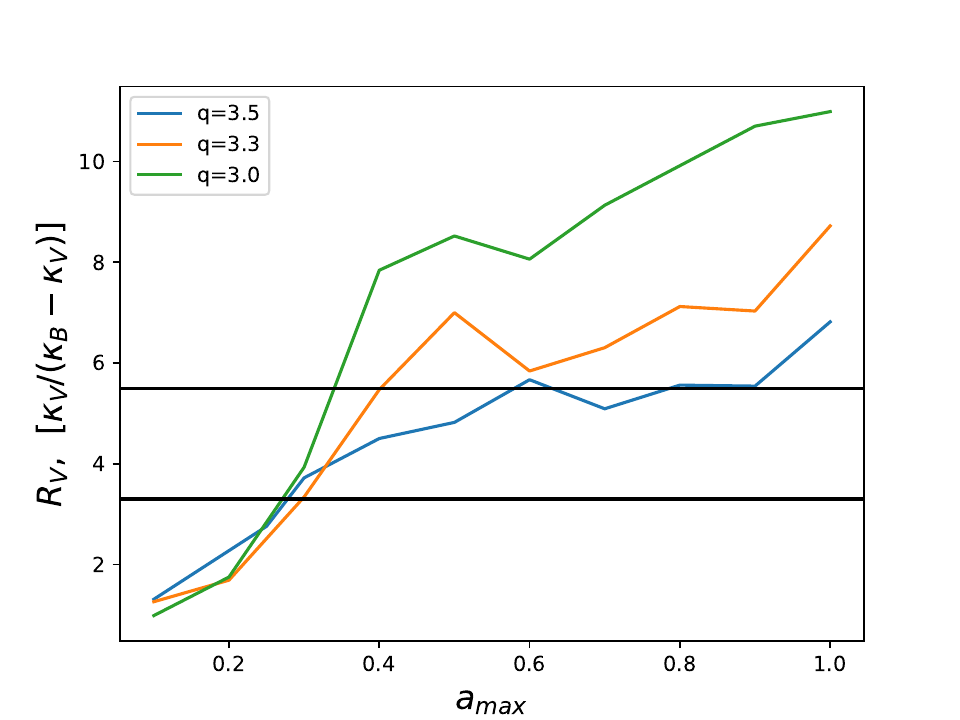}
    \caption{The reddening associated with maximum grain sizes for various power laws of the grain size distribution. The horizontal lines correspond to ${\rm R}_{\rm V} = 3.3$ and ${\rm R}_{\rm V} = 5.5$. }
    \label{fig:RvAmax}
\end{figure}

The cause of higher reddening is somewhat degenerate, but can be influenced by having larger maximum grain sizes, a change in the power law of the grain size distribution and the grain composition. We approximate ${\rm R}_{\rm V}$ from model dust distributions with different maximum grain size and power law of distribution using the opacities at the B and V band centres ${\rm R}_{\rm V} \approx \kappa_V/(\kappa_B-\kappa_V)$, which is plotted as a function of maximum grain size in Figure \ref{fig:RvAmax}. We explore two higher ${\rm R}_{\rm V}$ scenarios in our simple radiative transfer model. 

In the first higher ${\rm R}_{\rm V}$ scenario, the maximum grain size stays relatively small at $0.3\,\mu$m (meaning little scattering from the ambient medium) and the power law of the dust distribution is flattened to $q=3$ to give a higher ${\rm R}_{\rm V}$. As illustrated in the upper panel of Figure \ref{fig:HigherRV}, in that case we can still reproduce the flux drop in JWST filters at uniform densities of order $10^5$cm$^{-3}$. In our second ${\rm R}_{\rm V}=5.5$ scenario we keep a $q=3.5$ power law distribution but increase the maximum grain size in the ambient medium to $0.6\mu$m (see Figure \ref{fig:RvAmax}). In this case scattering from the ambient medium becomes important, and so in the absence of a full scattering radiative transfer calculation we are restricted to comparing the absorption opacities only. As illustrated in the lower panel of Figure \ref{fig:HigherRV}, the lack of distinct scattering from the trail means that higher, but plausible, densities in the range $10^6-10^7$\,cm$^{-3}$ are required to reproduce the contrast between the trail and ambient medium in that second higher ${\rm R}_{\rm V}$ case. 

For reference, the optical depths to the trail from the observer for a maximum grain size of 0.6\,$\mu$m and $q=3.5$ are $\sim 3.8$ at 1.15\,$\mu$m, $\sim 1.5$ at 1.4\,$\mu$m, $\sim 1.1$ at 1.82\,$\mu$m, and $\sim 0.73$ at 3\,$\mu$m. 

\begin{figure}
    \centering
    \includegraphics[width=0.99\columnwidth]{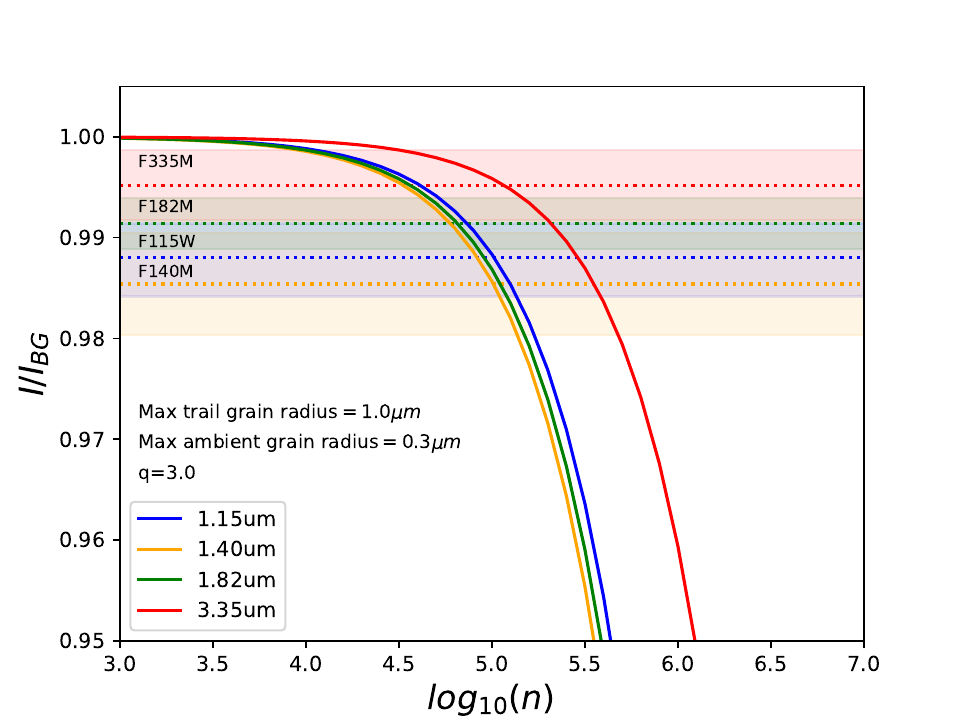}
    \includegraphics[width=0.99\columnwidth]{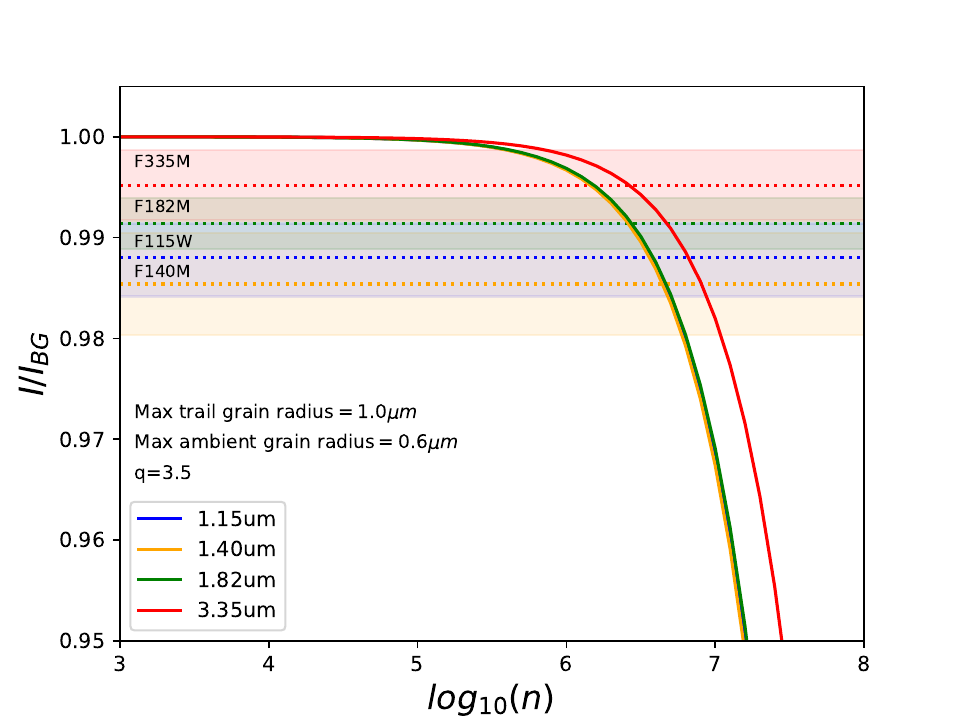}    
    \caption{The ratio of trail to ambient intensities, similar to Figure \ref{fig:Iratios}. In this are two scenarios where the ambient dust distribution gives a reddening ${\rm R}_{\rm V}\approx5.5$. In the upper panel the ambient medium has $a_{\textrm{max}}=0.3\,\mu$m and $q=3.0$. For a trail $a_{\textrm{max}}=1\,\mu$m and uniform medium we can still reproduce the scaling of intensity drops at $n\sim2\times10^5$\,cm$^{-3}$. The lower panel achieves ${\rm R}_{\rm V}\approx5.5$ by holding $q=3.5$ but increasing the maximum grain size in the ambient medium to 0.6$\mu$m. {Note that in the lower panel the 1.15\,$\mu$m line lies beneath the 1.4/1.82\,$\mu$m lines.}   }
    \label{fig:HigherRV}
\end{figure}



\subsection{Summary of radiative transfer modelling on the plausibility of observing the trail at high $A_V$}

{We have introduced a simple radiative transfer model with the objective of demonstrating that a trail physically associated with a brown dwarf at $A_V=52$ can give the contrast between trail and ambient medium observed by JWST NIRCam (see section \ref{sec:introductin2701954}). We find that this can be achieved at densities in the Dark Bay of $\sim10^{5}$\,cm$^{-3}$ in a uniform density medium. This requires no gas density enhancement in the trail itself, and can be achieved purely with a larger maximum grain size in the trail due to the strong sensitivity of the scattering opacity to maximum grain size in the NIR. 

However, there is a limit to how large that maximum grain size in the trail can be. The model works well for maximum grain size cases in the trail of 0.6 and 1$\,\mu$m, but for $10\,\mu$m dust in the trail cannot produce the relative contrast in different filters at any density because the opacity distribution flattens. 

Introducing a gas enhancement scales down the required  density of the Dark Bay. Based on arguments about the required extent of the cloud to give the observed extinction, as well as cloud stability, we favour a scenario where the trail gas density is at most only weakly enhanced relative to the surroundings, and that the cloud and trail gas density is high. If the ${\rm R}_{\rm V}=5.5$  in the Trapezium Cluster were due to larger dust grains ($\sim0.6$\,$\mu$m), then reproducing the contrast between trail and ambient medium would require higher densities, $10^6-10^7$\,cm$^{-3}$.  } 

\section{Possible wind-based mass loss and dust entrainment mechanisms}
\label{sec:winddriving}
We now evaluate possible mechanisms for leaving a wake of larger maximum grain size in the trail in a high density medium that is consistent with the picture established from the radiative transfer modelling. We begin by considering the various ways in which disc winds might be responsible.

\subsection{Dust entrainment in internal winds?}
The delivery and entrainment of dust in internal photoevaporative winds (driven by the host star) and magneto-centrifugal winds has been the subject of a number of studies \citep[e.g.,][]{2019ApJ...882...33G, 2021MNRAS.501.1127H, 2021MNRAS.502.1569B, 2023ASPC..534..567P}. In particular \cite{2021MNRAS.501.1127H} and  \cite{2021MNRAS.502.1569B} studied the actual delivery of dust from the disc to the base of photoevaporative winds, finding that the difficulty in delivering grains to the wind implies that dust entrained is always sub-micron in size and so would probably be insufficient to explain an enhancement in maximum grain size responsible for the trail of 270-1954. \cite{2019ApJ...882...33G} studied the entrainment of dust in magneto-centrifugal winds, finding it is typically sub-micron too in the case of a solar-type star. The critical maximum entrained grain size does have a $M_*^{-1/2}$ dependency \citep{2021MNRAS.502.1569B} and so in the case of the candidate brown dwarf studied here could lead to a factor $\sim 7$ increase, but that factor neglects the low UV/X-ray luminosity of such a low-mass object to drive the internal wind in the first place. Given these results, we expect that it is unlikely that internal winds are the source of micron-sized grain enhancements in the trail. 

\subsection{External photoevaporation due to X-ray superflares from V2445 Ori?}
270-1954 is in close projected separation to V2445\,Ori (7356\,au at 390\,pc, see Figure \ref{fig:Overview}), which is known to undergo extreme X-ray flaring events. For example, this includes a recent 30 hour X-ray flare with luminosity $2\times10^{32}$\,erg\,s$^{-1}$ \citep{2021ApJ...920..154G}, which is around an order of magnitude higher than the typical X-ray luminosity of similar mass ($\sim2$\,M$_\odot$) stars \citep{2021A&A...648A.121F} and around two orders of magnitude higher than fiducial X-ray luminosities considered when modelling internally driven X-ray photoevaporative winds \citep{2019MNRAS.487..691P, 2022MNRAS.514..535S}. This raises the possibility that flaring might externally heat circumstellar material around 270-1954 sufficiently to lead to mass loss and dust entrainment.

For a first estimate, we use the simple X-ray ionisation parameter--temperature prescription first developed by \cite{2010MNRAS.401.1415O, 2012MNRAS.422.1880O}. The ionisation parameter is
\begin{equation}
    \xi = \frac{L_\textrm{x}}{nD^2}
    \label{equn:ionparam}
\end{equation}
where $L_\textrm{x}$ is the X-ray luminosity of the source, $n$ is the local gas number density, and $D$ is the distance of the gas being considered from the X-ray source. In this scheme, that ionisation parameter is mapped onto a pre-computed temperature function and the corresponding X-ray heated temperature is used wherever the column between the X-ray source and the gas parcel is $\leq10^{22}$\,cm$^{-3}$. \cite{2019MNRAS.487..691P} developed column-dependent versions of the heating \citep[see also][]{2024arXiv240800848S} but for a first estimate we will use the simpler version. The temperature prescription is shown in Figure \ref{fig:xray} and the function used to generate it is given in Appendix~\ref{sec:xrayprescription}. 

\begin{figure}
    \centering
    \includegraphics[width=\columnwidth]{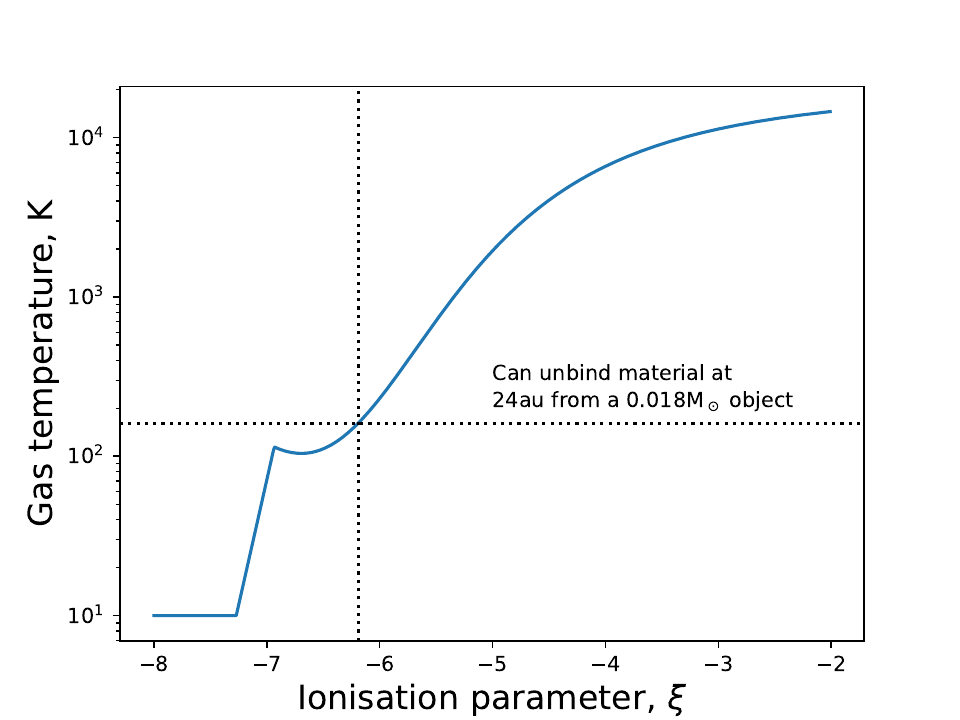}
    \caption{Temperature as a function of X-ray ionisation parameter as introduced by \protect\cite{2010MNRAS.401.1415O, 2012MNRAS.422.1880O}. Marked is the temperature and corresponding ionisation parameter above which material would be unbound at 24\,au from an 0.018\,M$_\odot$ object. }
    \label{fig:xray}
\end{figure}

To first order, we can estimate whether heating will unbind material using the gravitational radius. This tells us, at a given distance $R_g$ from a point mass $M_*$, at what temperature $T$ the sound speed exceeds the escape velocity
\begin{equation}
 T = \frac{GM_* \mu m_\textrm{H}}{k_B R_g }. 
 \label{equn:Rgrav_T}
\end{equation}
where $\mu m_\textrm{H}$ is the mean particle mass and $G$ is the gravitational constant. 

From Equation~\ref{equn:Rgrav_T}, to liberate material from a 24\,au molecular hydrogen disc around a 0.018\,M$_\odot$ central mass, we would thus require a temperature 160.5\,K\@. That temperature and corresponding minimum ionisation parameter ($\log_{10}\xi\sim-6.2$) is marked in Figure~\ref{fig:xray}.  We find that these X-ray flares can only heat the tenuous outer layers of the disc. For example, assuming that the distance $D$ is the projected separation between 270-1954 and V2445\,Ori and that $L_X=2\times10^{32}$\,erg\,s$^{-1}$ as in the recent flare, we can use Equation~\ref{equn:ionparam} to calculate the density of gas that could be warmed to that required 160.5\,K temperature as only $n=2.6\times10^4$\,cm$^{-3}$. This also assumes and requires that the column to V2445\,Ori is $\leq10^{22}$\,cm$^{-2}$.  Such a density is  below the sort of densities required for the basic radiative transfer model in Section~\ref{sec:basicRT}. This calculation is also based on an extremely strong and relatively short flare, and the frequency of those flares from V2445\,Ori is unknown. Overall, we therefore conclude that external X-ray irradiation by V2445\,Ori is unlikely to be responsible for driving the kind of mass loss that could lead to the trail. 

\subsection{FUV heating by $\theta^1$\,Ori\,C and/or $\theta^2$\,Ori\,A}
270-1954 lies at a projected separation from both $\theta^1$\,Ori\,C and $\theta^2$\,Ori\,A of around 0.5\,pc. Geometrically diluting the radiation field from $\theta^1$\,Ori\,C at that distance gives an FUV radiation field strength of order $10^4$\,G$_0$, which would certainly be sufficient to drive an external photoevaporative wind. Of course we know that there is high extinction along the line-of-sight between 270-1954 and the observer, but that does not mean the extinction is necessarily so high between 270-1954 and $\theta^1$\,Ori\,C, as illustrated in Figure \ref{fig:overviewCartoon}.

At only 0.018\,M$_\odot$, 270-1954 lies outside of the central star mass range of existing external photoevaporation calculations like the \textsc{fried} grid \citep{2023MNRAS.526.4315H}. We therefore run new bespoke 1D exernal photoevaporation calculations using the \textsc{torus-3dpdr} code \citep{2015MNRAS.454.2828B, 2019A&C....27...63H}, akin to those in \textsc{fried}. We use a 0.018\,M$_\odot$ central mass and run models for circumstellar discs of 5, 10, 15, and 20\,au in radius. We assume a temperature at 1\,au in the disc of 36\,K due to heating by the central star with a $T\propto R^{-1/2}$ profile. The surface density has an $R^{-1}$ profile with 100\,g\,cm$^{-2}$ at 1\,au.  We expose these discs to external FUV radiation fields ranging from $1-10^4$\,G$_0$, which corresponds to a range of extinction along the path connecting 270-1954 to the UV source from essentially zero to A$_{\rm V}\sim 8.5$\,mag. The geometry of the simulations is 1D, along the mid-plane through the disc and into the wind, since the mass loss is predominantly from the disc outer edge \citep[see][for full details]{2004ApJ...611..360A, 2023MNRAS.526.4315H}. 

Dust entrained in an external photoevaporative wind was studied by \cite{2016MNRAS.457.3593F}. The drag force in the wind can draw out dust up to some certain size $a_{\textrm{max}}$, which in the limit of the wind being very subsonic and ignoring centrifugal force on the grains themselves is approximated by
\begin{equation}
    a_{\textrm{max}} \approx \frac{\bar{v}}{GM_*}\frac{\dot{M}}{4\pi\mathcal{F}\bar{\rho}}
    \label{equn:amax}
\end{equation}
where $\bar{v}$ is the mean thermal velocity of the gas, $M_*$ the central star mass, $\dot{M}$ the mass loss rate, $\mathcal{F}=H_d/\sqrt{H_d^2 + R_d^2}$ the fraction of $4\pi$ steradians subtended by the disc outer edge for disc radius $R_d$ with scale height $H_d$, and $\bar{\rho}$ is the mean mass density of the dust grain material. 

\cite{2016MNRAS.457.3593F} demonstrated that because the maximum grain size is typically quite small (the exact value is dependent on the mass loss rate, but of order micron-sized), when grain growth has occurred in the disc, the amount of dust entrained in the wind and hence the extinction to the disc is reduced, resulting in higher mass-loss rates. In our calculations, we use a mean cross section in the wind appropriate for a situation where the dust in the disc had not undergone major grain growth, the only implication of which for this work is that the external photoevaporation calculations will have lower mass-loss rates than if the outer disc were depleted of entrainable dust, making it the conservative choice. 

\begin{figure}
    \centering
    \includegraphics[width=\columnwidth]{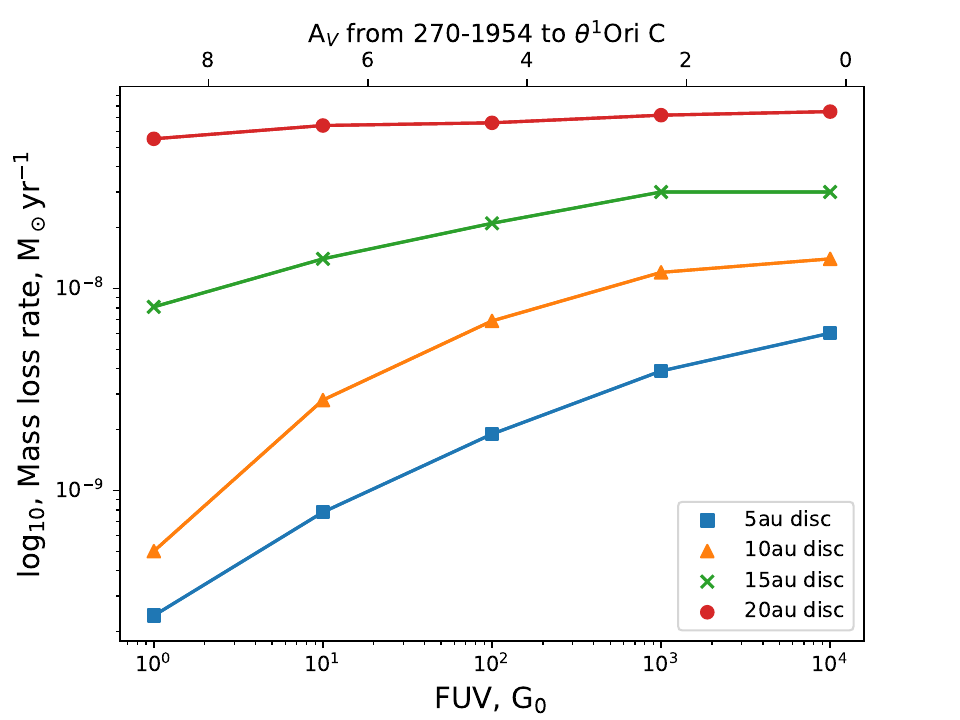}
    \includegraphics[width=\columnwidth]{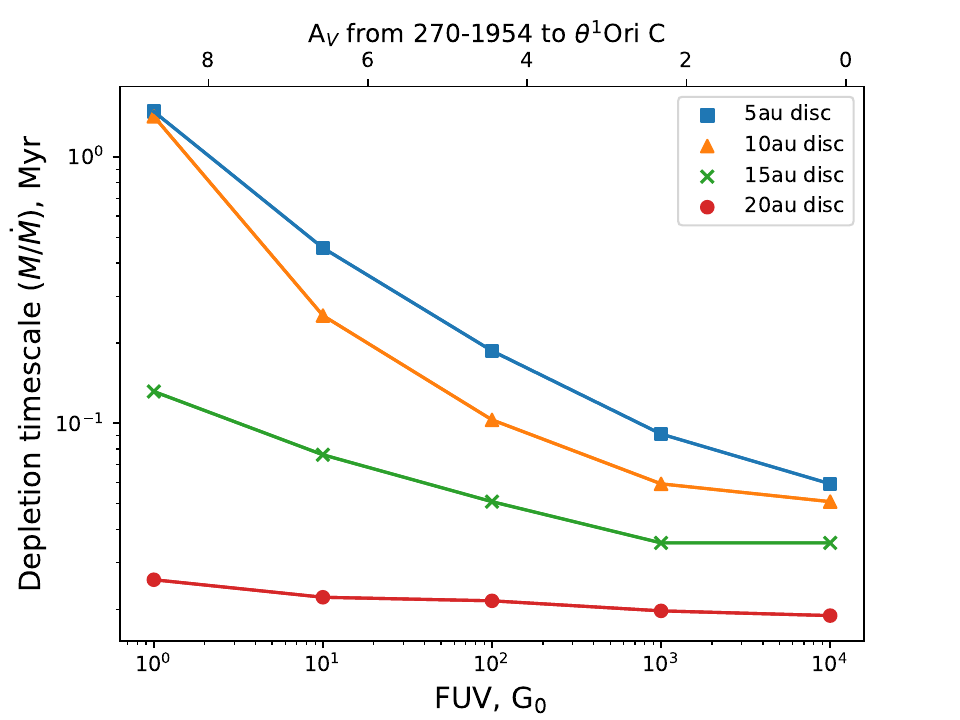}        
    \includegraphics[width=\columnwidth]{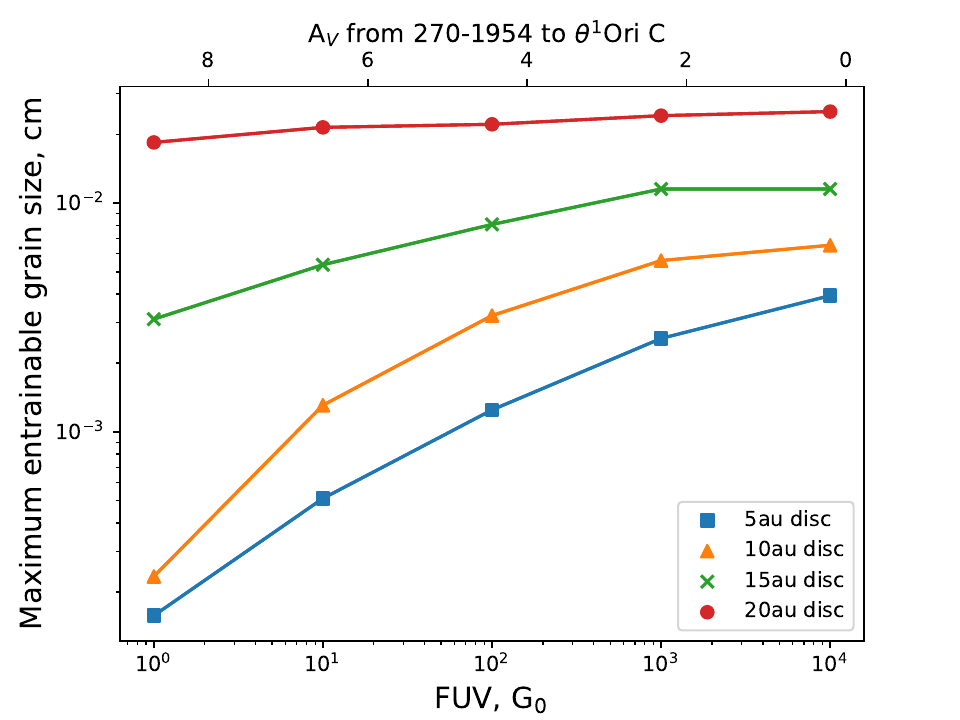}   

    \caption{{Mass-loss rates (top), depletion timescale (instantaneous ratio of disc mass to mass loss rate, middle) and maximum entrainable grain sizes in external photoevaporative winds (bottom) as a function of the external FUV radiation field strength. The different lines are for different disc radii. In all cases, it is possible to entrain micron-sized grains. For compact discs at low UV radiation field strength this can also take place associated with a long $\sim$Myr depletion timescale. Note that the depletion timescale of proplyds in the ONC is $<0.1$\,Myr.}} 
    \label{fig:FRIEDMdots}
\end{figure}

The resulting mass-loss rates, timescales to deplete the disc at that mass loss rate, and the corresponding predicted maximum entrainable grain size according to Equation~\ref{equn:amax} are given in Figure~\ref{fig:FRIEDMdots}. They demonstrate that a flow can be driven that entrains the required dust of order micron size to give the dark trail, even down to UV fields of at least as weak as $1$\,G$_0$ and for compact discs. Recall from section \ref{sec:basicRT} that 10\,$\mu$m dust in the trail was incompatible with the observed contrast. We therefore require the system to have a compact disc and be embedded. Taking the 1G$_0$ case, the corresponding $A_V\sim8.5$ to the UV source and $A_V\sim52$ to the observer would place 270-1954 at a depth $86$\,per cent through the total thickness of the Dark Bay along the line of sight. 270-1954 does therefore not need to reside on the far side of the bay in low density H\,\textsc{ii} region. Thus it seems plausible that the O~stars in the region could be driving the required mass loss to entrain the micron-sized dust required to give the contrast between the trail and surrounding nebula. 

The depletion timescale of the disc, shown in the middle panel of Figure~\ref{fig:FRIEDMdots} is short, but comparable to or higher than the depletion timescale of the hundreds of proplyds in the inner Orion Nebula \citep[e.g.][find of order $0.01-0.1$\,Myr]{1999AJ....118.2350H}.  

The trail is not directed towards the UV sources, so if the external photoevaporation scenario were the correct one the trail would be being left in the wake of the star as it propagates, rather than be streaming away along the vector connecting the UV source and evaporating disc. This could in principle be tested with proper motion measurements of 270-1954, however in practice the extreme faintness of this object means that JWST may be the only facility able to achieve this and a measureable proper motion for typical velocities in the region of $v\sim2$\,km\,s$^{-1}$ \citep{2022ApJ...926..141T} may take decades -- beyond the lifetime of JWST. 

\subsubsection{The width of the trail if resulting from a wind}
\label{sec:trailwidth}
We can also make a rough estimate of the width of the trail if it results from a wind. Key points of uncertainty for any such calculation are the velocities of source of the wind and the wind itself. If the source, i.e.\ 270-1954 in this case, moves rapidly, we can treat the geometry as cylindrical, otherwise a spherical treatment is required. External photoevaporative winds in low-UV environments usually have a launching and inner region that is subsonic, before becoming supersonic at some larger distance in the flow. Where the flow is subsonic, thermal pressure is important, but once it becomes supersonic, we can consider momentum-conserving transport. 

First we consider the scenario in which the brown dwarf has sufficiently high velocity, $v_{BD}$, \reply{such that the wind/trail geometry can be considered cylindrical sufficiently far behind the brown dwarf}. For a mass-loss rate, $\dot{M}$, and wind velocity, $v_w$, \reply{the radial momentum injected per unit length can be computed by integrating the wind momentum over solid angle to give $\frac{\pi}{4}\frac{\dot{M} v_w}{v_{BD}}$. Then, for a cold wind and a cold ISM in which thermal pressure can be neglected, the width of the tail may be approximated as a cylindrical `momentum-conserving snowplough' \citep{1996ApJ...459L..31W}, in which the momentum injected is equated with the momentum of an expanding, swept-up shell of ambient medium with density, $\rho_a$, and radius, $R$. This gives}
\begin{equation}
    \frac{\pi}{4}\frac{\dot{M} v_w}{v_{BD}} = \pi R^2\dot{R}\rho_a
\end{equation}
which leads to
\begin{equation}
    R = \left(\frac{3\dot{M}v_w}{4 \rho_a v_{BD}}\right)^{1/3}t^{1/3}.
\end{equation}
\reply{Here, we've assumed $R=0$ at $t=0$ for simplicity. Although $\dot{R}$ diverges at $t=0$ in this expression\footnote{An expression in which the shell velocity does not diverge can be found by using $R=\sqrt{3\dot{M}v_w/(4\pi\rho_a v_*^2)}\equiv R_0$ at $t=0$. For the parameters used in \autoref{eqn:wind_numbers}, $R_0\approx 20\,{\rm au}$. This approximation improves the solution significantly in the range $t\lesssim R_0/v_{\rm BD} \approx 50\,{\rm yr}$.}, the estimate reproduces exactly the asymptotic form of the solution for the tail far behind the brown dwarf \citep{1996ApJ...459L..31W}, as desired. Inserting typical values derived from the radiative transfer model, we find}
\begin{multline}
        R = 76.2\,\textrm{au} \left(\frac{t}{\textrm{kyr}}\right)^{1/3}\left(\frac{\dot{M}}{10^{-9}\,\textrm{M}_\odot\,\textrm{yr}^{-1}}\right)^{1/3}\left(\frac{\rho_a}{10^{-19}\,\textrm{g}\,\textrm{cm}^{-3}}\right)^{-1/3} \\
        \times \left(\frac{v_w}{0.2\,\textrm{km}\,\textrm{s}^{-1}}\right)^{1/3}\left(\frac{v_{BD}}{2\,\textrm{km}\,\textrm{s}^{-1}}\right)^{-1/3}. \label{eqn:wind_numbers}
\end{multline}
This applies until the ambient pressure terms become significant. The stellar velocity dispersion in the ONC is around 2\,km\,s$^{-1}$ and the escape velocity is $\sim 6.1$\,km\,s$^{-1}$ \citep[e.g.][]{2022ApJ...926..141T}. The timescale to cross the 1700\,au trail at 2\,km\,s$^{-1}$ is $\sim 4\,$kyr. However, the timescale for the expansion to become subsonic (assuming a supersonic wind velocity of 0.5\,km\,s$^{-1}$ and a sound speed of $0.2$\,km\,s$^{-1}$) is $840$\,yr. Taking $t=0.84$\,kyr, $\dot{M}=10^{-9}$\,M$_\odot$\,yr$^{-1}$, $v_w = 0.5\,$km\,s$^{-1}$ and $v_{BD}=2$\,km\,s$^{-1}$ this corresponds to a trail radius of $82, 38$ and 18\,au for number densities $10^5, 10^{6}$ and $10^{7}$\,cm$^{-3}$ respectively. The trail radius is $\sim24$\,au (2 pixels) at the end near the brown dwarf and broadens slightly to at least $\sim 36$\,au (3 pixels, 4 would increase the radius to 48\,au) towards the opposite end, so the size of the trail is broadly consistent with this simple model at plausible densities. The short timescale for ambient pressure terms to become important is also shorter than the likely timescale for the brown dwarf to traverse the trail based on the velocity dispersion in the cluster, which would explain the roughly uniform width of the trail.

\subsubsection{Summary of the external photoevaporative wind scenario}
In summary our simple arguments suggest that dust entrainment in an external photoevaporative wind is a plausible explanation for the dark trail. It can provide the maximum grain size enhancement in the trail required to give the trail-ambient contrast studied in the radiative transfer model. It can do so with 270-1954 embedded within the Dark Bay and is expected to give a fairly uniformly wide trail width of order the size observed.  

\section{A Bondi-Hoyle-Lyttleton accretion wake?}
\label{sec:BHL}
An alternative to a wind-driven source of material in the trail is a Bondi-Hoyle-Lyttleton type wake \citep[e.g.][]{1939PCPS...34..405H, 1944MNRAS.104..273B, 2004NewAR..48..843E, 2015PTEP.2015k3E01M}. Bondi-Hoyle accretion has been suggested by theoretical models as a potential means of driving disc evolution \citep{2024arXiv240507334P, 2024arXiv240508451W}. A Bondi-Hoyle-Lyttleton flow applies to a point mass accreting while propagating through the ambient medium. Material drawn into the trail of the object within some stagnation point 
\begin{equation}
    R_{\textrm{stag}} = \frac{2GM_{BD}}{v_{BD}^2}
\end{equation}
is accreted on to the object, while material drawn into the trail beyond that radius is left in the wake {(see Figure \ref{fig:BHL})}. Here $M_{\textrm{BD}}$ and $v_{\textrm{BD}}$ are the brown dwarf mass and velocity respectively. For the mass of 270-1954 that we estimated above this gives
\begin{equation}
    R_{\textrm{stag}}  \approx 8\,\textrm{au}\left(\frac{v_\textrm{BD}}{2\,\textrm{km}\,\textrm{s}^{-1}}\right)^{-2}.
\end{equation}
Thus the expectation is that if the 270-1954 trail were a Bondi-Hoyle-Lyttleton wake, then the accreting component would likely be unresolved by JWST and the trail would be the non-accreting wake. We note that with a gas sound speed of only $\sim 0.2$\,km\,s$^{-1}$, the brown dwarf velocity is therefore likely supersonic given the velocity dispersion of around 2\,km\,s$^{-1}$ \citep{2022ApJ...926..141T}. 

\begin{figure}
    \centering
    \includegraphics[width=0.6\columnwidth]{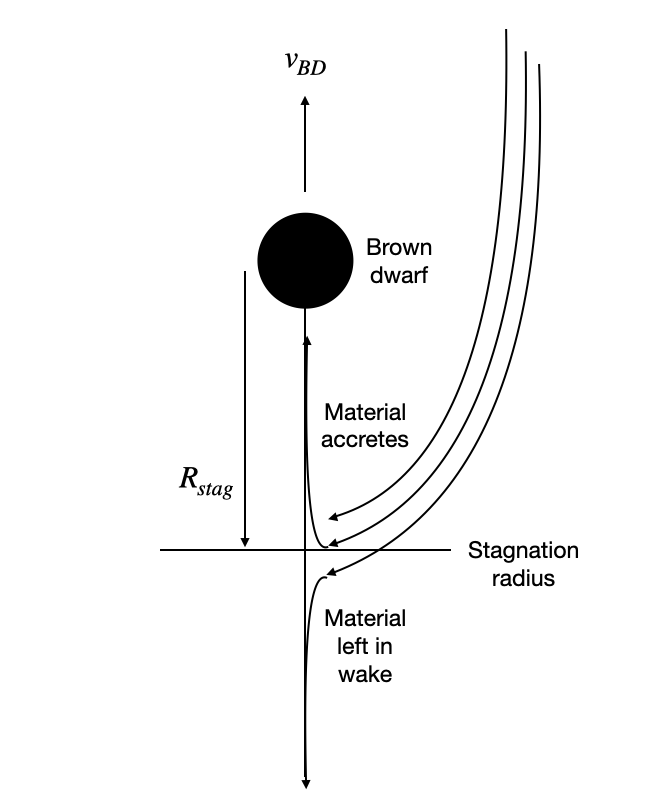}
    \caption{A schematic of the Bondi-Hoyle-Lyttleton accretion process. The brown dwarf propagates through the medium at velocity $v_{BD}$. Material drawn to the trailing mid-plane within the stagnation radius is accreted, while material beyond the stagnation radius is left in a wake which could then correspond to the 270-1954 trail. }
    \label{fig:BHL}
\end{figure}

Numerical simulations of such wakes show them to be complicated. They are subject to various possible instabilities, including the flip-flop instability, in which the wake swings from side-to-side with increasing amplitude \citep{1988ApJ...335..862F, 2005A&A...435..397F, 2008AJ....135.2380T, 2009ApJ...700...95B}. However, the nature of this is dependent on resolution and numerical method. The main goal of such simulations has been to study the accretion onto the central object and, to our knowledge, none exist that study the extended nature of the wake. But at the limit of the current observational data, the 270-1954 trail does not show signs of larger scale instability nor any significant increase in trail width with distance. Prior work has also not studied the properties of dust in the wake, which is likely going to be key for distinguishing the wind and wake scenarios. Future decoupled gas-dust simulations of Bondi-Hoyle-Lyttleton wakes will be necessary to achieve this. 

{We make an initial evaluation of the trail properties following \cite{1944MNRAS.104..273B} and \cite{2004NewAR..48..843E}, though this is known to not necessarily be representative of simulations as discussed above. Nevertheless, from \cite{2004NewAR..48..843E}, the mass per unit length in the wake is }
\begin{equation}
    m = \frac{2\pi G^2M_{BD}^2\rho_{a}}{v_{BD}^4}    
\end{equation}
which we convert to a number density assuming that the wake geometry is constant density and {a cylinder with the observed trail radius, $R_{\rm trail}=24$\,au, giving  }
\begin{equation}
    n_{\rm trail} = \frac{2\pi G^2M_{BD}^2 \rho_a }{v_{BD}^4}\frac{1}{\mu m_H}\frac{1}{\pi R_{\rm trail}^2}
    \label{equn:BHLtraildensity}
\end{equation}
{where $\mu$ is the mean molecular weight. We plot this as a function of the ambient number density and velocity in Figure \ref{fig:BHL_density}, which shows the expected behavior that lower velocities through denser media leads to higher wake densities. According to equation \ref{equn:BHLtraildensity} the density enhancement in the trail relative to the ambient medium depends upon the velocity through the medium and trail radius. We plot that enhancement of the density relative to the surroundings in Figure \ref{fig:BHL_enhancement}.  So in this simple approximation to the complicated reality expected from simulations, a density enhancement at the level of a factor 2 or greater would only be expected for velocities $<1\,$km\,s$^{-1}$, though the exact value would increase for a higher point mass or narrower trail radius. We note that the free fall time of a 1700\,au long, 42\,au radius cylinder \citep{2012ApJ...744..190T} is likely much longer than the crossing timescale of the trail by the brown dwarf. For example, the free-fall timescale is only faster for $v_{BD}<0.25$\,km\,s$^{-1}$ and $v_{BD}<0.9$\,km\,s$^{-1}$ for trail densities of 10$^8$\,cm$^{-3}$ and 10$^9$\,cm$^{-3}$ respectively.

\subsection{Summary of the BHL wake scenario}

{Overall our preliminary analysis above suggests that a BHL wake would likely be consistent with the picture of the trail being predominantly a maximum grain size enhancement, rather than the wake being a substantial gas density enhancement. This could  possibly be due to material stripped from a circum-brown dwarf disc, for example, \cite{2009ApJ...707..268M} studied the BHL accretion process for a star-disc system in a gas-only model, finding disc material is transferred into the wake.  Our calculations above also suggests that such a trail could survive for longer than the trail crossing time without undergoing global gravitational collapse. However BHL wakes can be unstable and so numerical models with decoupled dust-gas dynamics studying the nature of the extended wake are required.} 

Evidence for possible examples of continued infall from the ambient medium onto YSOs in the form of ``streamers'' has been observed in low-mass star-forming regions \citep[e.g.][]{2020NatAs...4.1158P, 2024A&A...683A.133G, 2024A&A...687A..71V}. If the trail associated with 270-1954 were to be confirmed as a Bondi-Holye-Lyttleton wake, it would be the first such example that we are aware of in a high-mass star-forming region.

\begin{figure}
    \centering
    \includegraphics[width=1.1\linewidth]{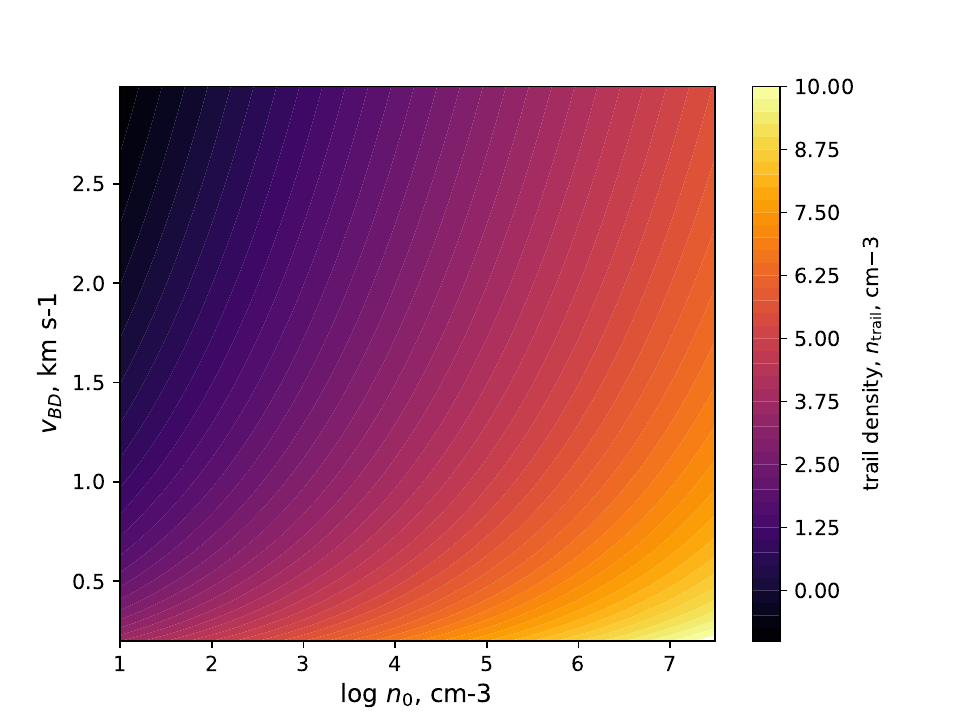}
    \caption{{Estimated number density in the BHL wake as a function of the number density in the ambient medium and velocity through the medium. Note that the velocities start at 0.2\,km\,s$^{-1}$, so we are always considering the supersonic regime for 10\,K gas.  }}
    \label{fig:BHL_density}
\end{figure}

\begin{figure}
    \centering
    \includegraphics[width=\linewidth]{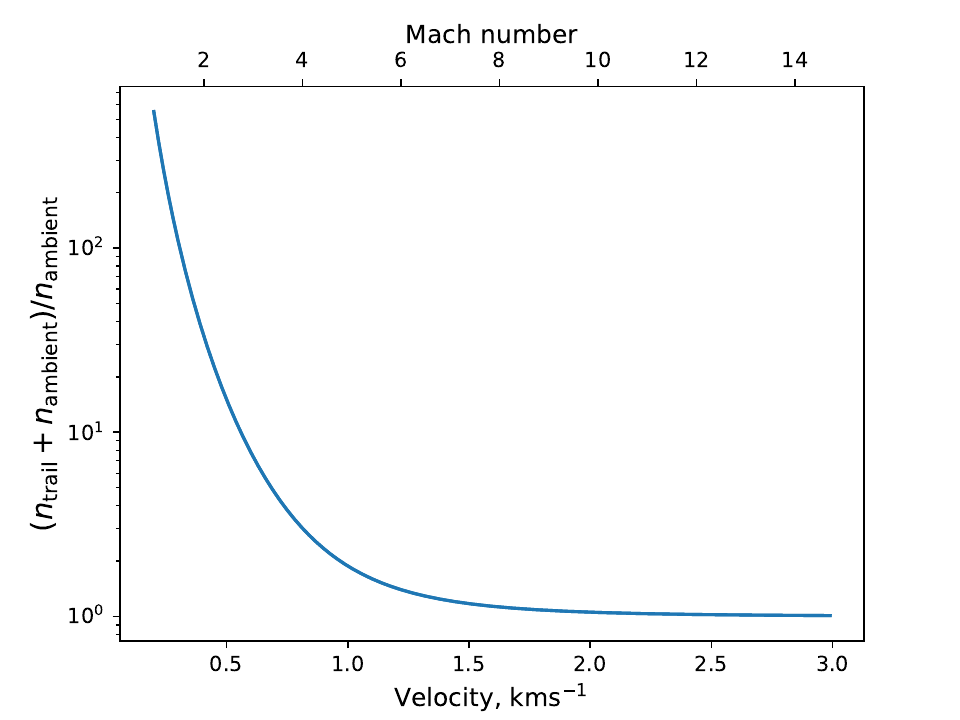}
    \caption{{The enhancement of density in the trail relative to the ambient medium as a function of the velocity of the point mass through the medium. Note that the velocities start at 0.2\,km\,s$^{-1}$, so we are always considering the supersonic regime for 10\,K gas.  }}
    \label{fig:BHL_enhancement}
\end{figure}

\section{Summary and conclusions}
From JWST observations of the Dark Bay in the inner Orion Nebula, we present a highly-extincted 0.018\,M$_\odot$ brown dwarf that appears to be coincident with one end of a very long, narrow dark trail, seen as a $\sim$1\% drop relative to the surrounding nebular intensity at $\sim1$\,$\mu$m. While we cannot fully prove a physical association between them at this stage, we use simple arguments to demonstrate that such a scenario is possible, both in terms of a trail being observed in the near-infrared at such high visual extinction, and in terms of there being mechanisms would could give rise to the trail.  

We demonstrate with a simple radiative transfer model that it is possible for the system to be observed through A$_{\rm V} = 52$\,mag of extinction and yield intensity contrasts between the trail and surrounding nebulosity consistent with the observations simply by introducing slightly larger dust grains in the trail. The primary argument is that the maximum grain size in the ISM is typically sub-micron, while material lost from a circum-brown dwarf disc could entrain micron-sized dust. At the $\sim$\,micron wavelengths observed by JWST, this then leads to order of magnitude changes in the scattering opacity due to the larger grain sizes and to the possible visibility of a trail.

We discuss four possible sources of the trail. Three mechanisms are due to disc winds from the brown dwarf system: internal winds (which are unlikely to entrain dust), X-ray driven winds from the nearby super X-ray flaring system V2445\,Ori, and attenuated FUV irradiation from Trapezium Cluster OB stars. We find that the X-ray driven irradiation does not seem plausible because the density of material that could be heated above the escape velocity is probably less than the density of the ambient medium. For the other two disc wind cases, we consider the scaling of the trail thickness with time, finding that a cylindrical momentum-driven trail expands weakly and would quickly be confined by external pressure, keeping the trail narrow. 

The other mechanism is a Bondi-Hoyle-Lyttleton wake. Little work has been done on the extended wakes of such accretion systems, though there is no evidence in the data for instabilities which are regularly seen in simulations of the process. It is thus difficult to assess the viability of this mechanism as the physical process responsible for the trail without future gas/dust dynamics simulations of the larger-scale extended wake. {However, we do make an initial estimate of the density in the trail in the BHL scenario, finding that it is consistent with the picture described above of being comparable to the ambient medium density and that the free fall time of the trail would be longer than the crossing timescale.}

For either disc wind or a Bondi-Hoyle-Lyttleton wake mechanisms, stronger evidence for the true physical association between the brown dwarf and the trail would come from future observations that measure the proper motion vector, which should be aligned along the trail vector. {However, at} typical stellar and brown dwarf velocities in the ONC, this would be a long-term project even with JWST\@. 


If the apparent association between 270-1954 and the trail can be physically proven, it would provide either direct evidence for the wider-reaching impact of external photoevaporation further from the centre of the Trapezium OB stars in the inner Orion Nebula, or potentially a unique view of Bondi-Hoyle-Lyttleton accretion in action. Either way, it would be a new and exciting example of environmental impact on star/planet formation. 

\section*{Acknowledgements}
We thank the anonymous referee for their thorough and insightful review of the paper. It significantly improved many aspects of the work. 

TJH acknowledges funding from a Royal Society Dorothy Hodgkin Fellowship and UKRI guaranteed funding for a Horizon Europe ERC consolidator grant (EP/Y024710/1). SGP acknowledges support through the ESA research fellowship programme at ESA's ESTEC, and would like to thank Victor See for helpful discussions and Katja Fahrion for valuable insights on the JWST calibration pipeline. RAB thanks the Royal Society for support through a University Research Fellowship.

The data presented in this paper were obtained with the Near Infrared Camera (NIRCam) on the NASA/ESA/CSA James Webb Space Telescope under Cycle~1 programme 1256, as part of the Guaranteed Time Observation allocation made to MJM upon selection as one of two ESA Interdisciplinary Scientists on the JWST Science Working Group (SWG) in response to NASA AO-01-OSS-05 issued in 2001.

\section*{Data Availability}

The JWST data here is all publicly available from the MAST science archive: \url{http://dx.doi.org/10.17909/vjys-x251}.



\bibliographystyle{mnras}
\bibliography{references} 





\appendix

\section{Forward scattering}
\label{sec:phasefunction}
Our simple radiative transfer model assumed negligible forward scattering from the trail into the trail. However, in the near-infrared, forward scattering can be significant. To estimate the possible impact we consider a \cite{1941ApJ....93...70H} scattering phase function 
\begin{equation}
    p(\theta) = \frac{1}{4\pi}\frac{1-g^2}{(1+g^2-2g\cos\theta)^{3/2}}
\end{equation}
where $\theta$ is the scattering angle and $1\leq g\leq 1$ is a parameter that controls the behaviour from pure back-scattering ($g=-1$), to isotropic scattering ($g=0$), and to pure forward scattering ($g=1$). We calculate the fraction of this function that scatters into the angular width of the trail (24\,au radius at 390\,pc), as shown in Figure \ref{fig:phasefunction}. In the near-infrared, the $g$ parameter is expected to get as high as around 0.7 and even in that case, the fraction of forward scattering into the trail is negligible, simply because the angular width of the trail is so narrow. 
\begin{figure}
    \centering
    \includegraphics[width=\columnwidth]{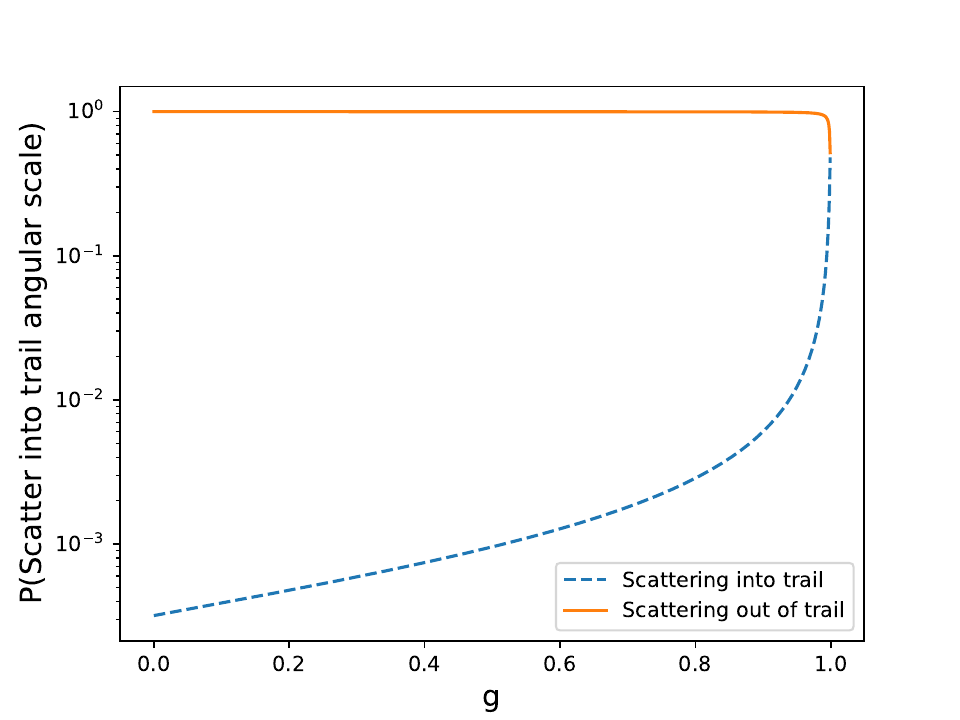}
    \caption{The fraction of a Henyey-Greenstein scattering phase function that would forward scatter from the trail into the trail. For the $g\leq0.7$ expected at $\sim1$\,$\mu$m, forward scattering into the trail is negligible given its tiny angular size.  }
    \label{fig:phasefunction}
\end{figure}

\section{X-ray ionisation parameter temperature function}
\label{sec:xrayprescription}
We calculate the X-ray heated gas temperature as a function of ionisation parameter $\xi$ using the prescription from \cite{2010MNRAS.401.1415O, 2012MNRAS.422.1880O}. The function is never fully written in those papers and so we include it here. The temperature is given by
$T = \max(T_{\textrm{hot}}, T_{\textrm{cold}}, 10.0)$. 
where
\begin{equation}
    T_{\textrm{hot}} =  10^{\left[(a_o\log_{10}\xi+b_o\log_{10}\xi^{-2})/(1+c_o\log_{10}\xi^{-1}+d_o\log_{10}\xi^{-2})+e_o\right]}
    \label{equn:Thot}
\end{equation}
and
\begin{equation}
    T_{\textrm{cold}} = 10^{(f_o\log_{10}\xi+g_o)}. 
    \label{equn:Tcold}    
\end{equation}
The numerical parameters are given in Table~\ref{tab:xrayparam}.

\begin{table}
    \centering
    \begin{tabular}{c|l}
    \hline
$a_o$ & \phantom{$-$}8936.2527959248299 \\ 
$b_o$ &           $-$4.0392424905367275 \\
$c_o$ & \phantom{$-$}12.870891083912458 \\
$d_o$ & \phantom{$-$}44.233310301789743 \\
$e_o$ & \phantom{$-$}4.3469496951396964 \\
$f_o$ & \phantom{$-$}3.15 \\
$g_o$ & \phantom{$-$}23.9 \\
\hline
    \end{tabular}
    \caption{The numerical parameters used in the X-ray heating prescription of equations \ref{equn:Thot}, \ref{equn:Tcold}. }
    \label{tab:xrayparam}
\end{table}

\section{Colour image of the system}

\begin{figure*}
    \centering
    \includegraphics[angle=0, width=\linewidth]{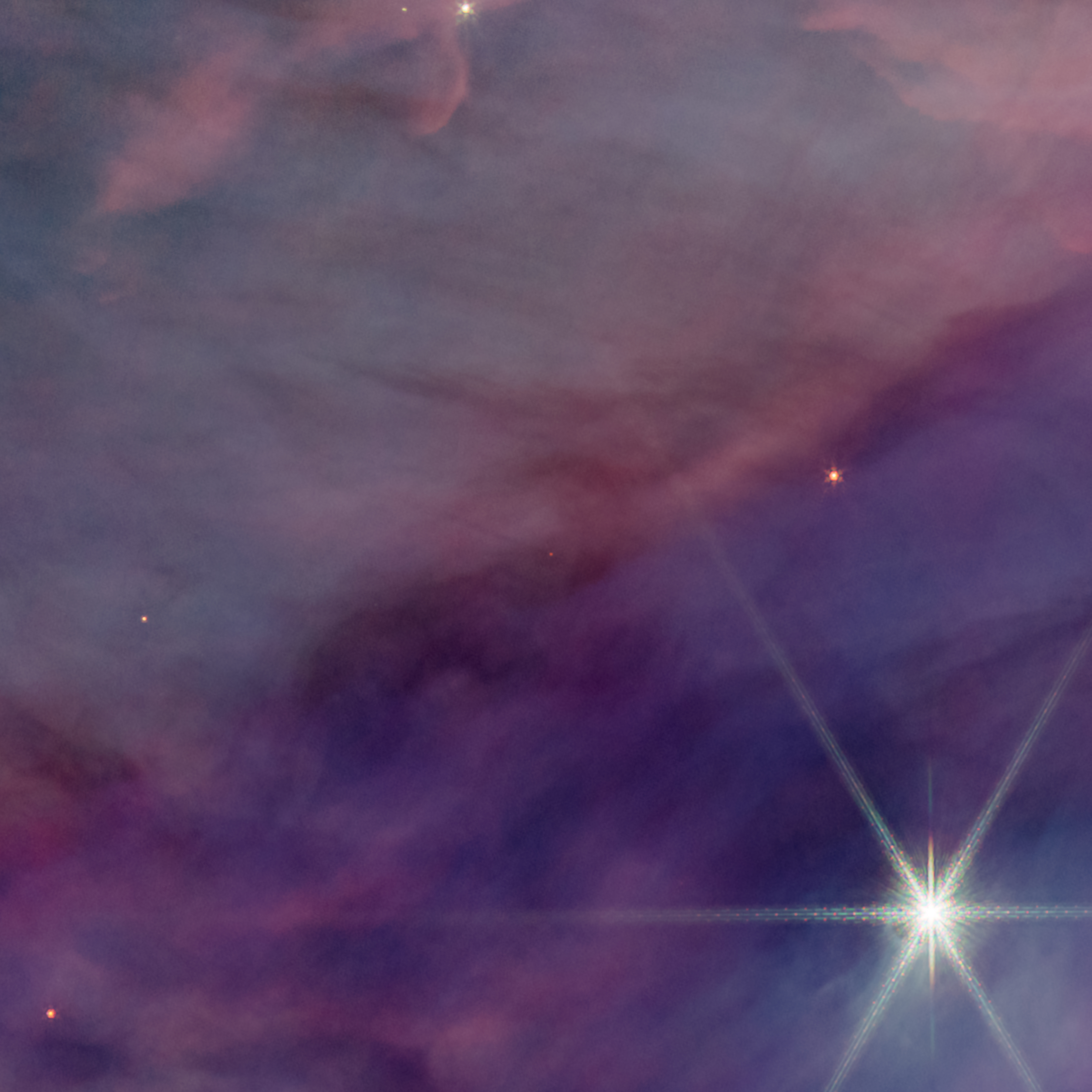}
    \caption{A 6 colour composite centred on 270-1954 in the F115W, F140M, F162M, F182M, F187N and F212N short wavelength filters. Note that north is down in this image.}
    \label{fig:colourimage}
\end{figure*}

\bsp	
\label{lastpage}
 \end{document}